\documentclass[aps,prd]{revtex4}
\usepackage{epsfig,graphics,graphicx}
\usepackage{slashed}
\usepackage[dvips]{feynmp}

\usepackage{color}

\newcommand{\bel}[1]{\begin{equation}\label{#1}}
\newcommand{\bal}[1]{\begin{eqnarray}\label{#1}}

\newcommand{\be}{\begin{equation}}
\newcommand{\ee}{\end{equation}}
\newcommand{\ba}{\begin{eqnarray}}
\newcommand{\ea}{\end{eqnarray}}

\newcommand{\bes}{\begin{equation*}}
\newcommand{\ees}{\end{equation*}}

\begin{document}

\def\be{\begin{eqnarray}}
\def\ee{\end{eqnarray}}
\def\dslash{\partial\!\!\!/}
\def\fd{_{fd}(E_k\,,\,\mu)}
\def\bpsi{\overline{\psi}}
\def\parag{\hspace* {0.18in }}
\def\nl{\newline}
\def\Ep{E_{{\bf p}}}
\def\Epu{E_{{\bf p}_1}}
\def\Epd{E_{{\bf p}_2}}
\def\Epi{E_{{\bf p}_i}}
\def\Eq{E_{{\bf q}}}
\def\sumint{\hbox{$\sum$}\!\!\!\!\!\!\!\int}
\def\LMS{\Lambda_{\overline{\textrm{MS}}}}

\title{Perturbative Yukawa theory at finite density: \\
the role of masses and renormalization group flow at two loops}

\author{ Let\'\i cia F. {\sc Palhares}\footnote{leticia@if.ufrj.br} and 
Eduardo S. {\sc Fraga}\footnote{fraga@if.ufrj.br}}
\affiliation{Instituto de F\'\i sica, 
Universidade Federal do Rio de Janeiro\\
C.P. 68528, Rio de Janeiro, RJ 21941-972, Brazil}


\begin{abstract}
Yukawa theory at vanishing temperature provides (one of the ingredients for) 
an effective description of the thermodynamics of a variety of cold and dense 
fermionic systems. We study the role of masses and the renormalization group 
flow in the calculation of the equation of state up to two loops within the 
$\overline{\textrm{MS}}$ scheme. Two-loop integrals are computed analytically 
for {\it arbitrary} fermion and scalar masses, and expressed in terms of well-known 
special functions. The dependence of the renormalization group flow on the number 
of fermion flavors is also discussed.

\vspace{0.5cm}


\end{abstract}

\maketitle

\section{Introduction}

Cold and dense fermionic systems are found in a wide range of physical 
environments and energy scales. In condensed matter, for instance, in the 
phenomena of antiferromagnetic ordering and superconductivity in the 
Hubbard model \cite{hubbard}. Increasing dramatically the density, Fermi 
pressure is responsible for compensating gravity in the hydrostatic equilibrium 
of compact stars, the Fermi gas being formed by electrons, neutrons or even 
quarks, depending on the density and formation process \cite{stars}. In nuclear 
and particle physics, hadrons undergo chiral and deconfinement transitions 
at sufficiently large baryonic densities  \cite{Stephanov:2007fk}, exhibiting a very 
rich phenomenology including several color superconducting phases 
according to the different possibilities for quark pairing \cite{Alford:2007xm}. 

Unfortunately, the thermodynamics of such plasmas in the case of vanishing temperature 
and finite density is not very amenable to first-principle calculations. On one hand, the 
so-called {\it sign problem} brings about major technical difficulties for performing Monte 
Carlo lattice simulations at finite chemical potential \cite{Laermann:2003cv}. 
On the other hand, although the perturbative series for the thermodynamic potential 
at zero temperature and finite density seems to be much better behaved than its counterpart 
at finite temperature \cite{Fraga:2001id,tony,andersen,Braaten:2002wi}, the values of 
chemical potential that are phenomenologically interesting usually have some overlap 
with a region where the coupling becomes large due to renormalization group running, 
so that perturbative calculations break down. Therefore, one is usually compelled to 
resort to effective theories, either to simplify the original description in terms of a 
fundamental theory, but still keeping its relevant symmetries, or to complement a 
given perturbative approach in the region where the coupling becomes large.

In spite of the fact that an effective theory does not require renormalizability to be well posed, 
this attribute is highly desirable if one is interested in investigating the behavior of physical 
quantities as the energy scale is modified, as discussed above, which can be accomplished 
via renormalization group methods. The requirement of renormalizability restricts 
considerably the spectrum of possibilities, and in most cases the effective theory will contain 
a Yukawa sector. In fact, the Yukawa theory is very common in particle and nuclear physics, 
not only because it is very convenient in building effective models but also due to 
the need for a mechanism of spontaneous symmetry breaking and mass generation 
in gauge theories. In condensed matter systems exhibiting phase transitions, such as 
the antiferromagnetic-superconductor transition mentioned above, the situation is 
analogous. Thus, perturbative Yukawa theory in a medium  has a wide range of applications. 
When it represents a sector of a fundamental theory, such as the quark-Higgs sector of 
the electroweak theory, one must assure that the scales are such that the Landau pole is not 
reached within the region of investigation, so that one can compute loop corrections and implement 
a renormalization group (RG) analysis. On the other hand, when it plays the role of part of an 
effective theory, one can avoid the Landau pole by simply introducing a cutoff beyond which the 
theory is meaningless. In the latter, one can choose to take effects from RG flow into account 
or not, depending on the system under investigation. 

In either case, the perturbative treatment of the Yukawa theory at finite density may be of help 
also for testing lattice simulations in the limit of small coupling. In fact, numerical studies at 
zero temperature and density of the Yukawa theory with a real scalar field \cite{Lee:1989xq} 
and of the chirally invariant Higgs-Yukawa model \cite{Gerhold:2007yb} have revealed a rich 
structure, in spite of the formal simplicity of the theory. In some cases perturbative calculations 
may complement in an efficient way Monte Carlo simulations in the investigation of the 
phase structure of a given effective theory \cite{GN}, in others the effective model can use lattice 
data as input for its parameters \cite{parameters}.

In this paper we investigate the role of masses and the RG 
flow in the calculation of the equation of state of cold and dense Yukawa theory 
up to two loops within the $\overline{\textrm{MS}}$ scheme. Previously we have presented 
preliminary results on the two-loop correction for the pressure with one massive fermion 
flavor and a massless scalar field \cite{Palhares:2007zz} and the influence of the RG running 
on the equation of state for $N_{F}$ flavors \cite{ijmpe}. The interest in studying results 
for different numbers of flavors, such as $N_F = 4$, is motivated not only by the 
phenomenological interest in different physical systems but also by on-going studies on the 
lattice using Kogut-Susskind fermions \cite{taurines}. Here we present full results for 
the thermodynamic potential with arbitrary masses and number of fermion species, including 
the effects from RG flow of couplings and masses. The two-loop momentum integrals are 
computed analytically for {\it arbitrary} fermion and scalar masses, the final result being 
expressed in terms of well-known special functions. Renormalization is implemented in the 
standard fashion in the $\overline{\textrm{MS}}$ scheme. 

The paper is organized as follows. In Section II we show our results for the two-loop expansion 
of the thermodynamic potential of the Yukawa theory at finite density, and discuss the 
renormalization procedure. Fixed mass and coupling results are presented in Section III. 
In Section IV we study the coupling and mass RG flows and their influence on the pressure. 
Section V contains our conclusions and outlook. Several technical details are left for 
a final appendix.

\section{Two-loop thermodynamic potential}

In what follows, we consider a gas of $N_{F}$ flavors of massive spin-$1/2$ fermions 
whose interaction is mediated by a massive real scalar field, $\phi$, with an interaction 
term of the Yukawa type, so that the Lagrangian has the following general form:
\begin{equation}
\mathcal{L}_{Y}
= \mathcal{L}_{\psi}+
\mathcal{L}_{\phi}
+\mathcal{L}_{int} \, ,
\label{Lyukawa}
\end{equation}
where
\begin{eqnarray}
\mathcal{L}_{\psi} &=& \sum_{\alpha=1}^{N_F} \bpsi_{\alpha}\left( i\dslash -m \right)
\psi_{\alpha} \, ,
\\
\mathcal{L}_{\phi} &=& \frac{1}{2} (\partial_{\mu}\phi)(\partial^{\mu}\phi)-
\frac{1}{2} m_{\phi}^2 \phi^2
-\lambda_3\phi^3-\lambda\phi^4 \, ,
\\
\mathcal{L}_{int} &=& \sum_{\alpha=1}^{N_F} g~\bpsi_{\alpha}\psi_{\alpha}\phi \, . 
\label{Lint}
\end{eqnarray}
Here, $m$ and $m_{\phi}$ are the fermion and boson masses, respectively, assuming all the 
fermions have the same mass, for simplicity. The Yukawa coupling is represented by $g$; 
$\lambda_3$ and $\lambda$ are bosonic self-couplings allowed by renormalizability. 
The latter play no role in the thermodynamics up to this order, unless in the presence of a 
nonzero scalar condensate \cite{footnote1},
as in the case of a spontaneously broken symmetry, where the 
condensate contributes to the effective masses. In this paper we assume $\langle\phi\rangle =0$, 
leaving the treatment of spontaneous symmetry breaking, as in the linear sigma model, 
for a future publication \cite{LSM}.

Although we are interested in the limit in which the temperature $T$ vanishes, it is 
technically more convenient to work with the imaginary time formalism of finite-temperature 
field theory, where the time dimension is compactified and associated with the inverse 
temperature $\beta=1/T$ \cite{kapusta-gale}. At the end, we shall take the limit 
$T\rightarrow 0$. As is well-known, to be consistent with the spin-statistics theorem for 
bosons ($B$) and fermions ($F$), one has to impose, respectively,  the fields to be periodic 
or anti-periodic in the imaginary time $\tau$, so that only specific discrete Fourier modes 
are allowed. Therefore, as is customary in finite-temperature field theory, integrals over the 
zeroth four-momentum component are replaced by discrete sums over Matsubara frequencies, 
denoted by $\omega^B_n=2n\pi T$ and $\omega^F_n=(2n+1)\pi T$, with $n$ integer. 
Taking finite density effects into account amounts to incorporating the constraint of 
conservation of the fermion number which, in practice, is implemented by a shift 
in the zeroth component of the fermionic four-momentum 
$p^0=i\omega_n^F\mapsto p^0=i\omega_n^F+\mu$, $\mu$ being the chemical potential.

From the partition function written in terms of the euclidean action for Eq. (\ref{Lyukawa}), 
 $Z_{Y}(T,\mu) = Tr ~\exp(-S_Y)$, one derives the perturbative series for the thermodynamic 
 potential $\Omega_{Y}\equiv - (1/\beta V) \ln Z_Y$:
\begin{equation}
\Omega_{Y}=
-\frac{1}{\beta V} \ln Z_0-
\frac{1}{\beta V}\ln \left[ 1+\sum_{\ell=1}^{\infty} \frac{(-1)^\ell}{\ell !}
\langle S_{int}^{\ell} \rangle_0 \right],
\end{equation}
where $V$ is the volume of the system, $Z_0$ is the partition function 
of the free theory and $S_{int}$ represents the euclidean interaction action. 
Notice that Wick's theorem implies that only even powers in the above 
expansion survive, yielding a power series in $\alpha_Y\equiv g^2/4\pi$
\cite{footnote2}.
%
Omitting purely bosonic contributions (which are $\mu$-independent)
and the diagrams representing counterterms, the thermodynamic potential 
up to two loops is given, diagrammatically, by

%

\vspace{-0.3cm}

\begin{fmffile}{fmftese}
\begin{eqnarray}
\Omega_Y&=& 
\frac{1}{\beta V}~N_F~
\parbox{10mm}{
\begin{fmfgraph*}(35,35)\fmfkeep{bolhaquark}
\fmfpen{0.8thick}
\fmfleft{i} \fmfright{o}
\fmf{fermion,left,tension=.08}{i,o}
\fmf{fermion,left,tension=.08}{o,i}
\end{fmfgraph*}}
\quad
+ ~\frac{1}{2}~\frac{1}{\beta V}~N_F~~
\parbox{10mm}{
\begin{fmfgraph*}(35,35)\fmfkeep{exchange}
\fmfpen{0.8thick}
\fmfleft{i} \fmfright{o}
\fmf{fermion,left,tension=.08}{i,o,i}
\fmf{dashes,label.dist=0.1w}{i,o}
\fmfdot{i,o}
\end{fmfgraph*}}
~~~
+~O(\alpha_Y^{2}) \, , 
\label{OmegaY}
\end{eqnarray}
\end{fmffile}

\noindent where solid lines represent the fermions and dashed lines stand for the bosons. 
The first diagram corresponds to the free gas contribution \cite{kapusta-gale}

\vspace{-0.4cm}

%
\begin{fmffile}{fmf2}
\begin{eqnarray}
\parbox{10mm}{
\begin{fmfgraph*}(35,35)\fmfkeep{bolhaquark}
\fmfpen{0.8thick}
\fmfleft{i} \fmfright{o}
\fmf{fermion,left,tension=.08}{i,o}
\fmf{fermion,left,tension=.08}{o,i}
\end{fmfgraph*}}
\quad
&=&
-2V~\int \frac{d^3{\bf p}}{(2\pi)^3} \left[ 
\beta\Ep
+\ln\left( 1+e^{-\beta(\Ep-\mu)} \right)
+\ln\left( 1+e^{-\beta(\Ep+\mu)} \right)
\right]
\, ,
\end{eqnarray}
\end{fmffile}

\noindent with $E_{\bf p}=\left({\bf p}^{2}+m^{2}\right)^{1/2}$. 
The zero-point energy divergent term can be absorbed by a convenient 
redefinition of the zero of the thermodynamic potential, since it is independent 
of $T$ and $\mu$. The $O(\alpha_Y)$ correction is given by the exchange term 
\cite{kapusta-gale}:

\vspace{-0.4cm}


%
\begin{fmffile}{fmf3}
\begin{eqnarray}
\parbox{10mm}{
\begin{fmfgraph*}(35,35)\fmfkeep{exchange}
\fmfpen{0.8thick}
\fmfleft{i} \fmfright{o}
\fmf{fermion,left,tension=.08}{i,o,i}
\fmf{dashes,label.dist=0.1w}{i,o}
\fmfdot{i,o}
\end{fmfgraph*}}
\quad &=& 
\beta V~g^2~\sumint_{P_1,P_2,K} \textrm{Tr}\left[ \frac{(2\pi)^3\beta 
~\delta^{(4)}(K-P_1+P_2)}{(\slashed{P}_1-m)(m_{\phi}^2-K^2)(\slashed{P}_2-m)} \right]
\, , \label{excRF}
\end{eqnarray}
\end{fmffile}

\noindent where the trace is performed over the Dirac structure, and the $4$-momenta are given 
in terms of the Matsubara frequencies and the $3$-momenta for fermions, 
$P_i=\left( p_i^0=i\omega_{n_i}^{F}+\mu \, ,\,  {\bf p}_i\right)$, and bosons, 
$K=\left( k^0=i\omega_l^{B}\, ,\, {\bf k} \right)$. We use the metric tensor 
$g^{\mu\nu}=\textrm{diag}(+,-,-,-)$ and the following notation for the sum-integrals:
\be
\sumint_P &=&T\sum_{n}\int \frac{d^3{\bf p}}{(2\pi)^3} \, .
\ee

The calculation of the exchange diagram is very similar for different theories and is a 
standard exercise in finite-temperature field theory, the main difference in details 
coming from the tensor structure of each theory under investigation. Previous results for cold 
and dense systems can be found, for instance, in Ref. \cite{Furnstahl:1989wp} for 
nuclear matter, in Refs. \cite{Russos,Freedman:1976dm,Toimela:1984xy} for QED and in Refs. 
\cite{Kapusta:1979fh,Toimela:1984xy,Farhi:1984qu,Freedman:1976ub,Baluni:1977ms,Fraga:2004gz} 
for QCD (see also Ref. \cite{kapusta-gale}). After some long but straightforward algebra, 
we can write the exchange term for the Yukawa theory in the following form (see Appendix):

\vspace{-0.3cm}

%
\begin{fmffile}{fmf8}
\begin{eqnarray}
\parbox{10mm}{
\begin{fmfgraph*}(35,35)\fmfkeep{exchange}
\fmfpen{0.8thick}
\fmfleft{i} \fmfright{o}
\fmf{fermion,left,tension=.08}{i,o,i}
\fmf{dashes}{i,o}
\fmfdot{i,o}
\end{fmfgraph*}}
\quad &=& 
\beta V~g^2
\int \frac{d^3{\bf p}_1 d^3{\bf p}_2}{(2\pi)^6}
~\frac{1}{2\omega_{12}\Epu\Epd}\Bigg\{
\overline{\mathcal{J}}_{+}~\omega_{12}~\Sigma_1
+
\overline{\mathcal{J}}_{-}~\omega_{12}~\Sigma_2
+\nonumber \\
&&
+2\Big[ \overline{\mathcal{J}}_{-}~E_+-\overline{\mathcal{J}}_{+}~E_- \Big]~
n_b(\omega_{12})~N_f(1)
-\nonumber \\
&&
-\Big[ \overline{\mathcal{J}}_+ (E_-+\omega_{12})-\overline{\mathcal{J}}_- (E_+-\omega_{12}) \Big]~N_f(1)
-\nonumber \\
&&
-2~\overline{\mathcal{J}}_-~E_+~n_b(\omega_{12})
-\nonumber \\
&& 
-\overline{\mathcal{J}}_-(E_+-\omega_{12})
\Bigg\}
\, , \label{excRes}
\end{eqnarray}
\end{fmffile}

\vspace{-0.3cm}

\noindent where we have defined the following functions:
\be
\overline{\mathcal{J}}_{\pm} &\equiv& 
-2~\frac{m^2-{\bf p}_1\cdot {\bf p}_2\pm \Epu\Epd}{E_{\mp}^2-\omega_{12}^2}
= 1-\frac{4m^2-m_{\phi}^2}{E_{\mp}^2-\omega_{12}^2} \, ,
\label{barJpm}
\\
N_f(i) &\equiv& n_f(\Epi+\mu) +n_f(\Epi-\mu) \, ,
\label{Nfs}
\\  
\Sigma_1 &\equiv& n_f(\Epu+\mu)~n_f(\Epd+\mu)+n_f(\Epu-\mu)~n_f(\Epd-\mu) \, ,
\label{Sig1}
\\
\Sigma_2 &\equiv& n_f(\Epu+\mu)~n_f(\Epd-\mu)+n_f(\Epu-\mu)~n_f(\Epd+\mu) \, ,
\label{Sig2}
\ee
with $E_{\pm} \equiv \Epu \pm \Epd$ and 
$\omega_{12} \equiv \left( |{\bf p}_1-{\bf p}_2|^2+m_{\phi}^2\right)^{1/2}$;  
$n_{b}(\omega)=\left[\exp(\beta \omega)-1\right]^{-1}$ and $n_{f}(E)=\left[1+\exp(\beta E)\right]^{-1}$ 
are the Bose-Einstein and the Fermi-Dirac distributions, respectively.

The physical meaning of each term in Eq. (\ref{excRes}) is clear. The first two lines 
are quadratic in the statistic distributions, representing contributions coming from 
the scattering of particles from the medium. These are therefore ultraviolet finite due to the 
exponential suppression of the integrands implemented by the distributions. The 
other terms contain contributions coming from the scattering of virtual particles. The 
last term is a pure vacuum contribution, independent of $T$ and $\mu$, and can be 
absorbed by a redefinition of the zero of the thermodynamic potential. On the other 
hand, the remaining terms mix medium and vacuum particles, being linear in the 
statistic distributions, and call for renormalization.

The divergent contributions that are linear in $n_{b}$, ${\bf L}_b$, and linear in 
$n_{f}$, ${\bf L}_f$, can be written in terms of amputated ($amp$) vacuum ($vac$) 
self-energies evaluated on the mass shell ($m.s.$) as follows (see Appendix):
%

\vspace{-0.3cm}

\begin{fmffile}{fmf82}

\begin{eqnarray}
{\bf L}_f&=&-2~\beta V~{\left\{ {
\sumint_P~(-1)\textrm{Tr}\left[\frac{1}{\slashed{P}-m}
\left(i~
\parbox{23mm}{
\begin{fmfgraph*}(65,40)\fmfkeep{autoenergiaF}
\fmfpen{0.8thick}
\fmfset{arrow_len}{0.15w}
\fmfleft{i} \fmfright{o}
\fmf{fermion,tension=.8,label=$P_1$}{i,v1}
\fmf{fermion,tension=.5,label=$P_2$}{v1,v2}
\fmf{plain,tension=1.2}{v2,o}
\fmf{dashes,left,tension=0,label=$K$}{v1,v2}
\fmfdot{v1,v2}
\end{fmfgraph*}}
\right)_{\stackrel{amp}{m.s.}}^{vac}
\right]
}
\right\}_{matter}
}
\, , \label{LfCT}
\\
\nonumber \\
{\bf L}_b&=&-~\frac{\beta V}{N_F}~{\left\{ {
\sumint_Q \frac{1}{m_{\phi}^2-Q^2}
\left(i~
\parbox{23mm}{
\begin{fmfgraph*}(65,40)\fmfkeep{autoenergiaB}
\fmfpen{0.8thick}
\fmfleft{i} \fmfright{o}
\fmf{dashes,tension=.8,label=$K$}{i,v1}
\fmf{phantom,tension=.7}{v1,v2}
\fmf{fermion,left,tension=0,label=$P_1$}{v1,v2}
\fmf{fermion,left,tension=0,label=$P_2$}{v2,v1}
\fmf{dashes,tension=1}{v2,o}
\fmfdot{v1,v2}
\end{fmfgraph*}
}
\right)_{\stackrel{amp}{m.s.}}^{vac}
}
\right\}_{matter}
}
\, .
\label{LbCT}
\end{eqnarray}


\noindent Here, $matter$ means that the pure vacuum part has already been subtracted,
and, while $K=(\omega_{{\bf k}},{\bf k})$ and $P_1=(\Epu,{\bf p}_1)$ are evaluated on the 
mass shell, $Q=(i\omega_l^B,{\bf k})$ and $P=(i\omega_n^F+\mu,{\bf p}_1)$ are not.
Implementing 
the renormalization procedure for the self-energies above in the $\overline{\textrm{MS}}$ scheme, 
one obtains the renormalized expressions (see Appendix)
\begin{eqnarray}
{\bf L}_f^{\textrm{ren}}
&=&
-2~\beta V\frac{\alpha_Y}{4\pi} ~4m^2
\int \frac{d^3{\bf p}_1}{(2\pi)^3}~\left[ 
\frac{N_f(1)}{2\Epu}
\right] 
\left[ \alpha_1 \right]
\, , \label{LfCTRen}
\\ \nonumber \\
{\bf L}_b^{\textrm{ren}}
&=&
-2~\beta V~\frac{\alpha_Y}{4\pi}
\int \frac{d^3{\bf k}}{(2\pi)^3}
\left[ \frac{2n_b(\omega_{\bf k})}{2\omega_{\bf k}}
\right]~
~\left[ 2~\alpha_2+6~\alpha_3
\right]
\, , \label{LbCTRen}
\end{eqnarray}
where
\be
\alpha_1&=&
-4 \frac{m_{\phi}}{m}\left( 1-\frac{m_{\phi}^2}{4m^2} \right)^{\frac{3}{2}}
\left\{ \tan^{-1}\left[ \sqrt{\frac{1}{\frac{4m^2}{m_{\phi}^2}-1}} \right]
+\tan^{-1}\left[ \frac{\frac{1}{2}-\frac{m_{\phi}^2}{4m^2}}{\sqrt{ \frac{m_{\phi}^2}{4m^2} }\sqrt{1-\frac{
m_{\phi}^2}{4m^2}}} \right] \right\}
\nonumber \\
&& +\frac{7}{2}-\frac{m_{\phi}^2}{2m^2}-\frac{3}{2}\log\left( 
\frac{m^2}{\Lambda^2} \right)+\frac{m_{\phi}^2}{m^2}\left( \frac{3}{2}
-\frac{m_{\phi}^2}{4m^2} \right)
\log\left( \frac{m^2}{m_{\phi}^2} \right)
\, ,
\label{alpha1res}
\\
\alpha_2&=& m^2- \frac{1}{6}m_{\phi}^2
\, ,
\\
\alpha_3&=& \frac{2}{3}\left[ 2m^2-\frac{5}{12}m_{\phi}^2 \right]-
\frac{1}{3}m_{\phi}^2\left( \frac{4m^2}{m_{\phi}^2}-1 \right)^{\frac{3}{2}}
\tan^{-1}\left[ \frac{1}{\sqrt{\frac{4m^2}{m_{\phi}^2}-1}} \right]
-\left( m^2-\frac{m_{\phi}^2}{6} \right)\log\left( \frac{m^2}{\Lambda^2} \right)
\, ,\label{alpha3res}
\ee
and $\Lambda$ is the renormalization subtraction point.

In the limit of vanishing temperature and in the absence of a scalar condensate, 
the Bose-Einstein distribution vanishes, so that the 
purely bosonic diagrams, omitted in Eq. (\ref{OmegaY}), do not contribute. 
The Fermi-Dirac distribution simplifies to the Heaviside 
step function, $\lim_{T\to 0} n_f(\Ep -\mu)= \theta(\mu-\Ep)$, signaling the 
occupation of all states in the interior of the Fermi surface. Simplifying all the terms 
in the thermodynamic potential, we obtain in the cold and dense limit
\begin{eqnarray}
\Omega_Y &=&
- N_F~\frac{1}{24\pi^2}
\left[
2~\mu p_f^3-3 m^2~u
\right]-
\frac{1}{2}~N_F~4\pi\alpha_Y
\left[J_1
+ \frac{1}{16\pi^{4}} m^{2} u \alpha_{1}
\right] \, ,
\label{OmegaYResT0}
\end{eqnarray}
where $u\equiv\mu p_f-m^2\log\left( \frac{\mu+p_f}{m} \right)$, $p_{f}$ is the 
Fermi momentum $p_f \equiv \sqrt{\mu^2-m^2}$ and we have defined the 
integral
\be
J_1 \equiv -
\int \frac{d^3{\bf p}_1 d^3{\bf p}_2}{(2\pi)^6}
~\frac{1}{2\Epu\Epd}~
\overline{\mathcal{J}}_{+}~\theta(\mu-\Epu)~\theta(\mu-\Epd)
\label{defJ1}
\, .
\ee
The evaluation of the integral above when both fields are massive is highly non-trivial, 
and results are usually presented in the limit of at least one vanishing mass, which is the 
proper case in gauge theories, or numerically \cite{kapusta-gale}. We 
obtained the following complete analytic result for {\it arbitrary} values of $m$, $m_{\phi}$ 
and $\mu$ (see Appendix for details)
\be
J_1 &=& \frac{1}{(2\pi)^4} \left\{ 2m^2\left( 1-\frac{m_{\phi}^2}{4m^2} \right)
~\mathcal{I}_I-\frac{1}{2}~u^2 \right\}
\, ,
\label{ResJ1final}
\ee
where the function $\mathcal{I}_I $ is given by
\be
\mathcal{I}_I &=&
\mu^2~\frac{m_{\phi}^2}{2m^2} ~\log\left[ 
	\frac{m_{\phi}^2}{4p_f^2+m_{\phi}^2}
	\right]
	+
	\left(1-\frac{m_{\phi}^2}{2m^2}\right)~\frac{u^2-\mu^2 ~p_{f}^2}{m^2}
+\nonumber \\&&
+\frac{m_{\phi}}{m} \sqrt{1-\frac{m_{\phi}^2}{4m^2}}
~(\mu~p_f+u)~D_{\tan}
+ 
p_f^2 -2m^2~\overline{\mathcal{K}}_{log}\left(\frac{p_f}{\mu+m},\frac{m_{\phi}^2}{4m^2}\right)
\, ,
\ee
with
\be
D_{\tan}  &\equiv&
\tan^{-1}\left(
\frac{-p_{f} m_{\phi}}{2m(\mu+m)\sqrt{1-\frac{m_{\phi}^2}{4m^2}}}
\right) + 
\tan^{-1}\left(
\frac{p_{f}}{m_{\phi}\sqrt{1-\frac{m_{\phi}^{2}}{4m^{2}}}}
\left[2-\frac{m_{\phi}^2}{2m(\mu+m)} \right]
\right)
\, . \nonumber\\
\ee
For $\bar{x}\equiv p_f/(\mu+m)~<~1$ and $z\equiv m_{\phi}^2/4m^2$, 
we can write the function $\overline{\mathcal{K}}_{\log}(\bar{x},z)$ 
as
\be
\overline{\mathcal{K}}_{\log}(\bar{x},z) &=& \sqrt{z}~\Delta(\bar{x},z)+\frac{1}{2}\sqrt{z(z-1)}~
\mathcal{C}_{\textrm{Li}}(\bar{x},z)\, ,
\label{barKlogRES}
\ee
where
\begin{eqnarray}
\!\!\!\!\mathcal{C}_{\textrm{Li}}(\bar{x},z) &=& 
\Bigg[
\mathbf{Li}_{2}\left( 1-\left( z-\sqrt{z(z-1)} \right)(1-\bar x) \right)-
\mathbf{Li}_{2}\left( 1-\left( z+\sqrt{z(z-1)} \right)(1-\bar x) \right)+\quad\quad
\nonumber\\
&& \!\!\!\!\!\!\!\!\!\!\!\!\!\!\!\!\!\!\!\!\!\!\!\!\!\!\!\!\!\!\!\!\!\!+
\mathbf{Li}_{2}\left( \frac{1-\bar x}{1+\bar x}\left[
1-\left( z+\sqrt{z(z-1)} \right)(1-\bar x) \right] \right)-
\mathbf{Li}_{2}\left( \frac{1-\bar x}{1+\bar x}\left[
1-\left( z-\sqrt{z(z-1)} \right)(1-\bar x) \right]  \right)
\Bigg]
+\Bigg[ \bar x \mapsto -\bar x \Bigg]
\, , \label{Clif}
\end{eqnarray}
with the Dilogarithm Function \cite{Abramowitz} defined as 
$\mathbf{Li}_{2}(z) \equiv \sum_{n=1}^{\infty} (z^n/n^2)$. The quantity 
$\Delta$ is defined in two regions, according to the ratio $z$, as 
\be
\Delta(\bar{x},z) &=& \left\{ \begin{array}{ll} 
    \Delta_<(\bar{x},z) & \textrm{, if $z< 1$}\\ 
    \Delta_>(\bar{x},z) & \textrm{, if $z> 1$}
    \end{array} \right. \, ,
\ee
where
\be
\Delta_<(\bar{x},z) &=& \sqrt{1-z}\Bigg\{  
\log(1-\bar x)\Bigg[  
\tan^{-1}\left( \frac{\sqrt{1-z}}{\sqrt{z}} \right)-\tan^{-1}\left(  
\frac{\sqrt{z(1-z)}}{z+\left(\frac{1+\bar x^2}{2\bar x}- 1\right)^{-1}}
\right)
\Bigg]+\nonumber \\
&&
\!\!\!\!\!\!\!\!\!\!\!\!\!\!\!\!\!\!\!\!\!
+\log(1+\bar x) \Bigg[ \tan^{-1}\left( \frac{\sqrt{1-z}}{\sqrt{z}} \right)-\tan^{-1}\left(  
\frac{\sqrt{z(1-z)}}{z+\left(\frac{1+\bar x^2}{2\bar x}+ 1\right)^{-1}}
\right) 
-\pi\left[ 1-\theta\left( m z \frac{(1+\bar x)^2}{1-\bar x^2}-\frac{2m\bar x}{1-\bar x^2} \right) \right]
\Bigg]
\Bigg\}
\, ,\\
\Delta_>(\bar{x},z) &=&\frac{1}{2}\sqrt{z-1}\Bigg\{
\log\left( \frac{1-\bar x^2}{(1+\bar x)^2} \right)\log\left( \frac{|\sqrt{z}(1+\bar x^2)-
2\bar x\sqrt{z-1}|}{|\sqrt{z}(1+\bar x^2)+
2\bar x\sqrt{z-1}|} \right)+\nonumber\\&&+
\pi^2 \Bigg[ -\theta[z(1-\bar x^2)-1]-\theta\left(  
1+ \frac{\bar x}{(1-\bar x^2)\sqrt{z(z-1)}} [2-z(1-\bar x^2)]
\right)
+\nonumber\\&&
+\theta\left(  
1- \frac{\bar x}{(1-\bar x^2)\sqrt{z(z-1)}} [2-z(1-\bar x^2)]
\right)+1 \Bigg]  
\Bigg\}
\, .\label{ResJ1last}
\ee

With this general result for the thermodynamic potential for the cold and dense Yukawa 
theory, we can analyze the effects from the different nonzero masses and their competition, 
the role of the interaction and the renormalization scale, as well as the influence of the RG 
flow of the coupling and masses.

\section{Results at a fixed energy scale}

Let us start our study by the simpler case with fixed energy scale $\Lambda$.
In practice, one can choose an appropriate value for $\Lambda$, in this case a
parameter in a given effective theory, by imposing, e.g., known experimental
constraints for the system under consideration. In the next section, we investigate 
the role played by $\Lambda$ as the running scale in the RG flow of $g$, $m$ and 
$m_{\phi}$. For fixed $\Lambda$, the main issue is the influence of the masses which 
we illustrate separating the analysis into two cases, the first with $m_{\phi}=0$, and then the general one.

\subsection{Massless boson}

When $m_{\phi}=0$, the integrals involved in the computation of the thermodynamic
potential at finite density simplify dramatically. In fact, the former complicated
functions $\alpha_1$ and $J_1$ reduce to
\be
\lim_{m_{\phi}\to 0}~\alpha_1 &=& 
\frac{7}{2}-\frac{3}{2}\log\left( 
\frac{m^2}{\Lambda^2} \right)
\, ,
\ee
\be 
\lim_{m_{\phi}\to 0}~J_1 &=& 
\frac{1}{32\pi^4} \left\{ 3~u^2-4~p_f^4\right\}
\, ,
\ee
so that the thermodynamic potential assumes the much simpler form
\be
\Omega_Y &=&
- N_F~\frac{1}{24\pi^2}
\left[
2~\mu p_f^3-3 m^2~u
\right]
-N_F~\frac{\alpha_Y}{16\pi^3}
\left\{
3~u^2-4~p_f^4
+
m^2~
u~
\left[ 7-3\log\left( 
\frac{m^2}{\Lambda^2} \right) \right]
\right\}
\, .
\ee
It is interesting to notice that the inclusion of mass for the fermions
brings the presence of logarithmic corrections one order down in $\alpha_Y$.
This is a general feature, also manifest in theories such as QCD. In massless
QCD, for instance, one has $\sim \alpha_s$ corrections at two loops and
$\sim \{\alpha_s^2,\alpha_s^2\log \alpha_s,\alpha_s^2\log(\Lambda/\mu)\}$
at three loops \cite{Fraga:2001id,tony}. However, in the massive case one
finds not only $\sim \alpha_s$ terms, but also a contribution $\sim m^2\alpha_s\log(\Lambda/m)$
at two loops \cite{Fraga:2004gz}, analogous to the one that can be seen above. In our case,
this feature will be important when we incorporate the RG running of
$\alpha_Y$ and $m$. On the other hand, since the explicit dependence on $\Lambda$
is logarithmic, the effect of its variation without the RG flow
does not affect significantly the pressure \cite{Palhares:2007zz} unless 
for very large values of the coupling, where perturbation theory is meaningless.

\begin{figure}[htb]
\begin{minipage}[t]{75mm}
\includegraphics[width=8cm]{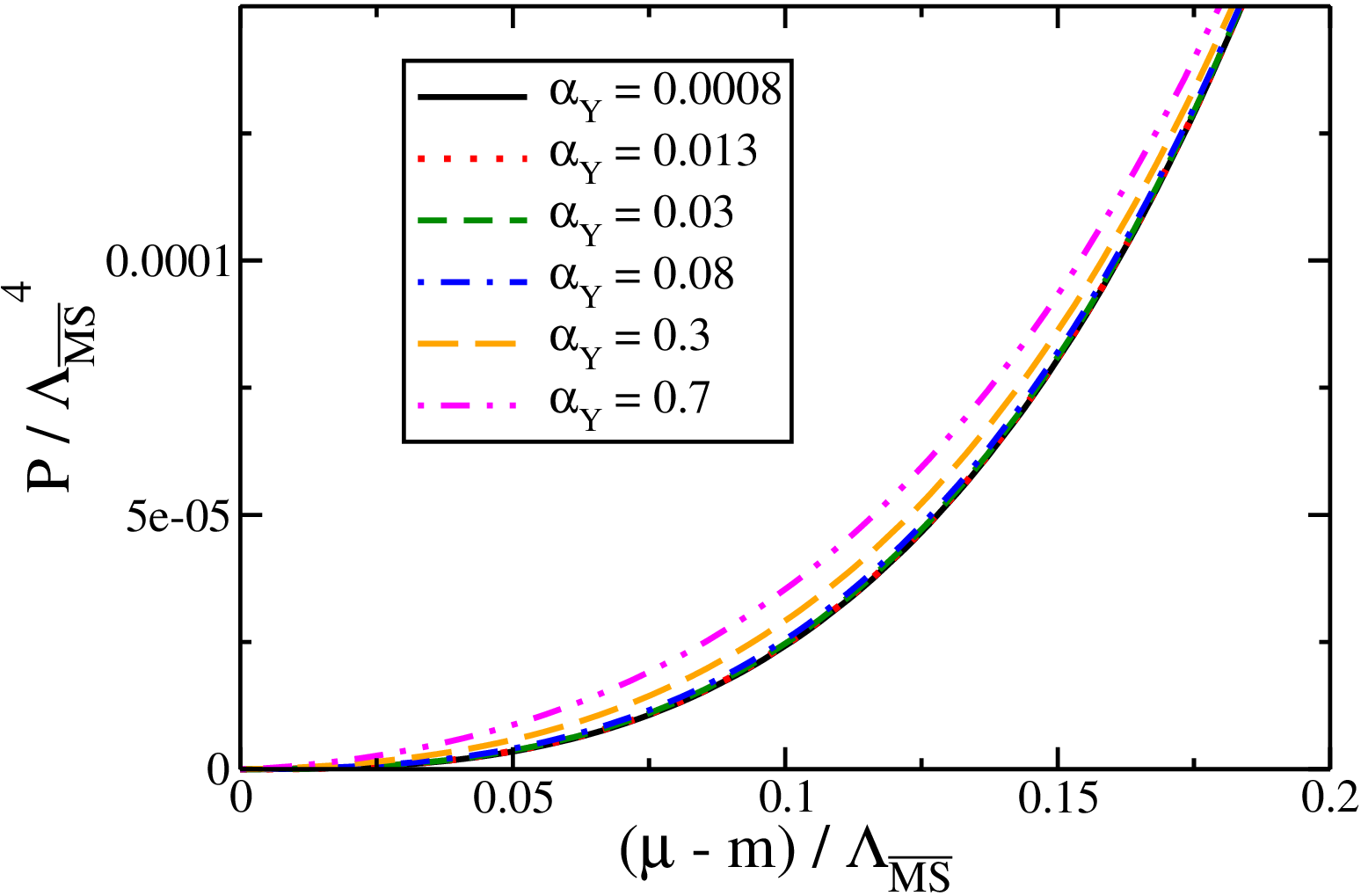}
\caption{Pressure normalized by the scale $\Lambda=\LMS$ as 
a function of the fermion chemical potential for different values of 
the coupling $g$. The fermion mass is fixed at $m=0.1\LMS$.}
\label{Pvsmumenosm-varg}
\end{minipage}
\hspace{2cm}
\begin{minipage}[t]{75mm}
\includegraphics[width=7.7cm]{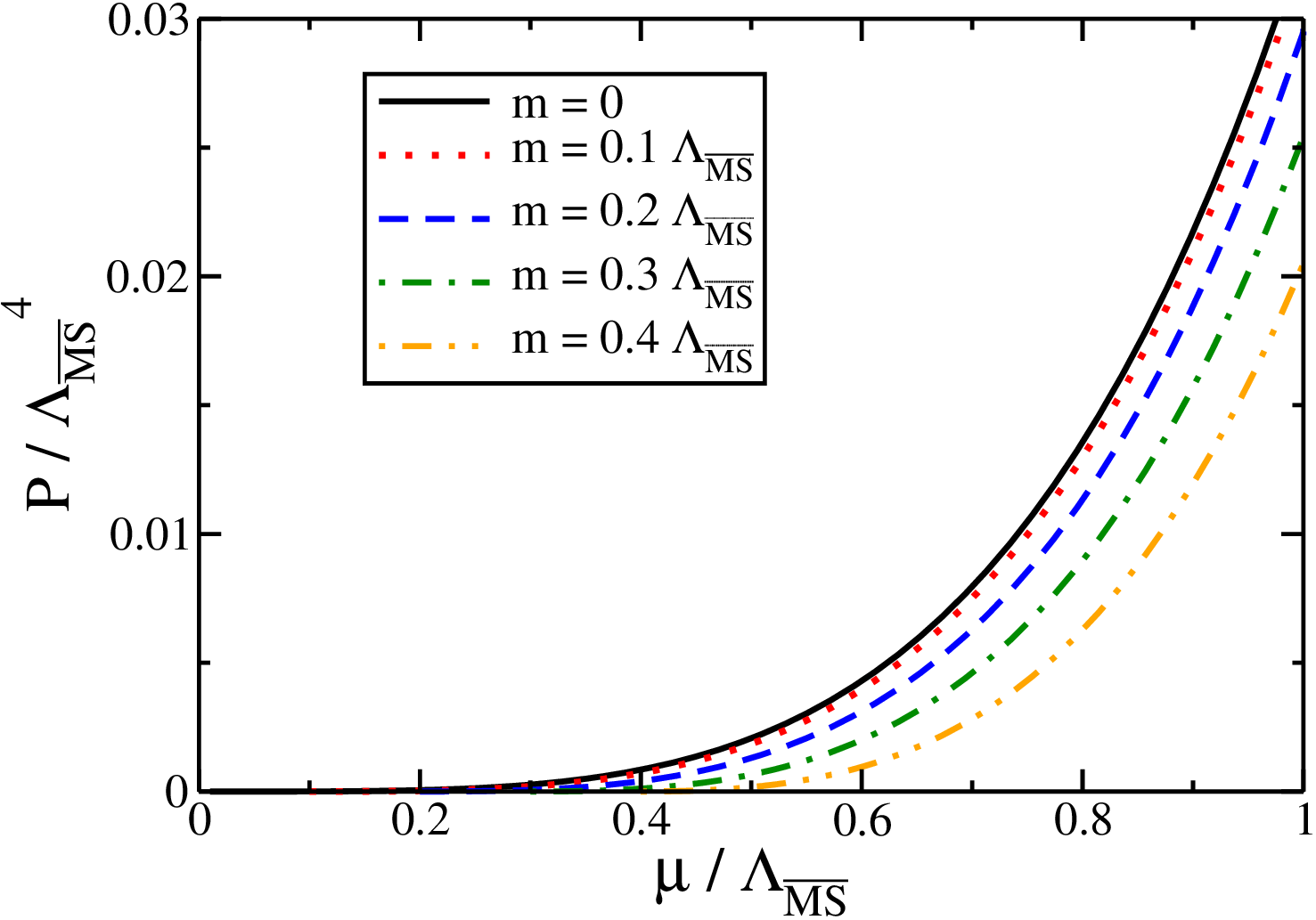}
\caption{Pressure normalized by the scale $\Lambda=\LMS$ as 
a function of the fermion chemical potential for different values of 
the fermion mass $m$. The coupling is fixed at $\alpha_Y=1/4\pi$.}
\label{Pvsmu-varm}
\end{minipage}
\end{figure}

Figs. \ref{Pvsmumenosm-varg} and \ref{Pvsmu-varm} illustrate, respectively,
interaction and fermion mass effects on the two-loop pressure, $P=-\Omega_Y$,
as a function of the chemical potential $\mu$ in units of the reference scale $\Lambda=\LMS$.
Fig. \ref{Pvsmumenosm-varg} shows that the pressure is raised as the system 
considered interacts more strongly, reaching a variation of $\sim 50\%$ 
in comparison with the free case for $\alpha_Y\approx 0.72$ and $(\mu-m)=0.1~\LMS$.
This behavior observed at small chemical potentials is not maintained 
at large scales: there is a crossing at $(\mu-m)\approx 0.25~\LMS$ so 
that the interaction effects tend to reduce the pressure of highly dense media,
although the corrections in this regime appear to be less significant ($\lesssim 20\%$).
In fact, since the Yukawa coupling is treated perturbatively,
it is quite reasonable that interaction effects do not modify drastically the
thermodynamics within the domain of validity of our calculations.
Nevertheless, depending on the specificities of the system under investigation,
corrections of a few percent as the ones obtained above may have impact on
physical predictions.


%

On the other hand, the fermion mass is not linked directly to the 
expansion parameter and its effects are therefore not expected to be constrained by
the perturbative framework.
Indeed, Fig. \ref{Pvsmu-varm} shows that the influence of finite 
fermion masses on the thermodynamics can be sensibly more consequential: 
at $\mu=0.6\LMS$, for instance,
the inclusion of fermions with mass $m=0.4\LMS$ lowers the pressure by a factor
of $\sim 1/5$ in comparison with the massless case. Even for masses one order of magnitude
smaller than the reference scale $\LMS$, the corrections are sizable for sufficiently
small chemical potentials. Therefore, finite fermion masses alter significantly 
the thermodynamics of the Yukawa theory in a wide range of the parameter space
and approximations that neglect them should be implemented cautiously.

This result represents another indication of the potential importance of fermion mass effects. 
Recently, the modifications brought about by finite masses have received increasing attention in
different contexts, either because they have been underestimated before or due to
the interest in high-precision tests, in experiments and in more realistic lattice simulations.
In calculations within the Standard Model, for instance, there is a whole literature on
multi-loop integrals that is progressively turning its attention to the inclusion of contributions 
due to non-vanishing fermion masses \cite{grozin-book}. In QCD it was also pointed out recently 
that finite quark masses, especially the strange quark mass that is not that small compared to 
typical scales in QCD, should play an important role in the critical region of the chiral and 
the deconfining phase transitions \cite{Stephanov:2007fk}, possibly bringing relevant 
astrophysical consequences \cite{Fraga:2004gz}.
These results for the Yukawa theory signal that also within effective theories mass effects
bring relevant corrections.

\subsection{General Case}

\begin{figure}[htb]
\vspace{.5cm}
\begin{minipage}[t]{75mm}
\hspace{-1cm}
\includegraphics[width=7.5cm]{Pvsmphi.eps}
\caption{Pressure normalized by the fermion mass $m$ as a function 
of the ratio $m_{\phi}/m$ for $\mu=4m/3$ and $\Lambda=10~m$.}
\label{Pvsmphi}
\end{minipage}
\hspace{1cm}
\begin{minipage}[t]{85mm}
\vspace{-7.7cm}
\hspace{-3cm}
\hspace{0.3cm}
\includegraphics[width=11cm]{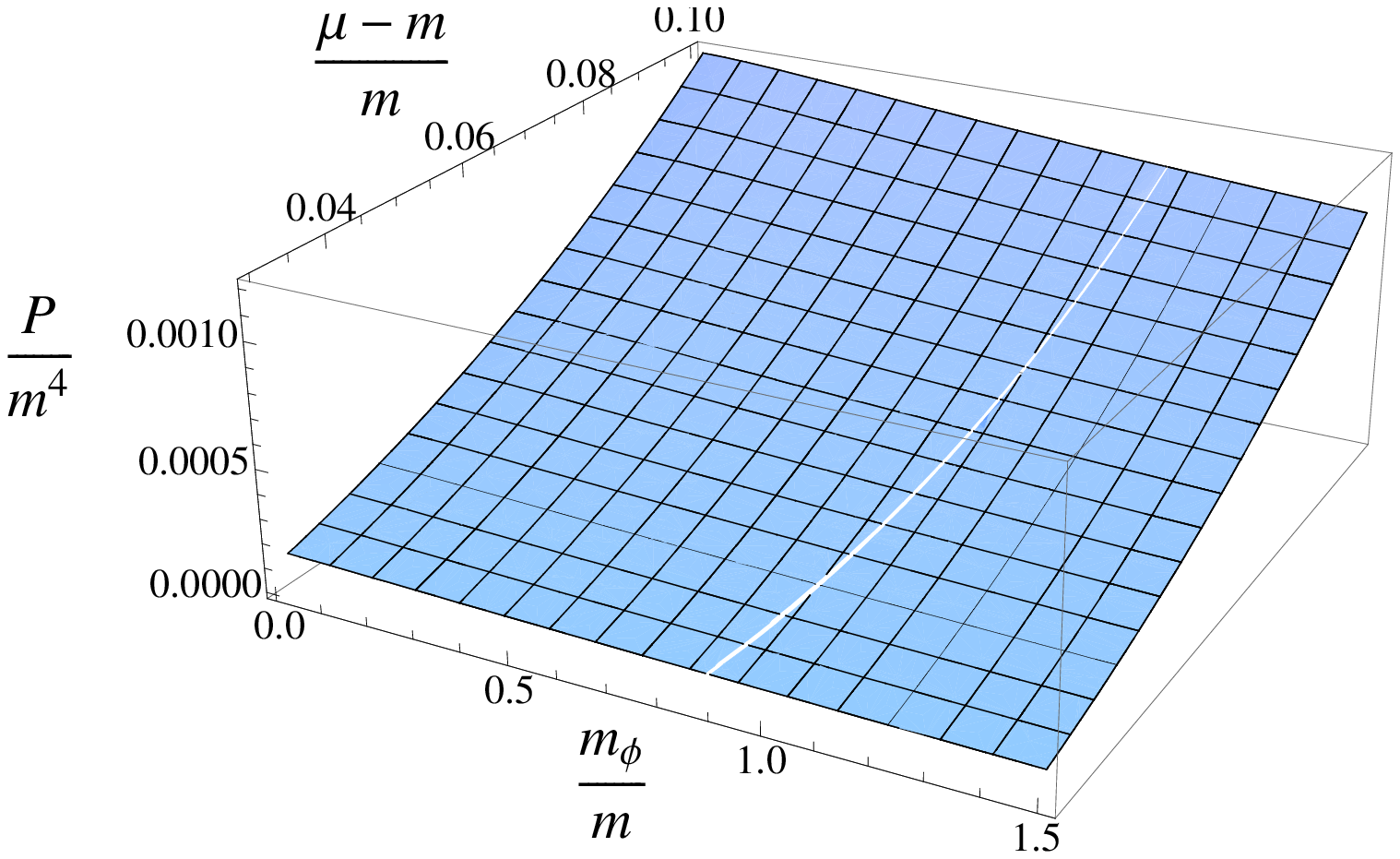}
\hspace{-1.5cm}
\vspace{-6.7cm}
\caption{Pressure normalized by the fermion mass $m$ as a function 
of the ratio $m_{\phi}/m$ and $\mu$ ($\Lambda=10~m$).}
\label{3dgraf.ps}
\end{minipage}
\end{figure}

Let us now consider the full massive case. The influence of the boson mass $m_{\phi}$
on the thermodynamics of the Yukawa theory is shown in Fig. \ref{Pvsmphi}.
Since we are analyzing the cold and dense regime, the pressure investigated
consists essentially in a quantum Fermi pressure. In this vein, it is convenient
to interpret the results in terms of a free quasi-particle theory: the variation of 
the pressure can be seen as a consequence of the modification of the effective mass
of the quanta present in the system through the radiative corrections
to the fermionic self-energy due to the coupling with the bosons.
We showed above, in Fig. \ref{Pvsmu-varm}, that for a given chemical potential
the pressure for a theory of heavy fermions is lower than the one for light fermions.
Therefore, Fig. \ref{Pvsmphi} indicates that for $m_{\phi}/m$ sufficiently small
the self-energy is negative, diminishing the fermionic effective mass and increasing
the pressure. For the case illustrated in Fig. \ref{Pvsmphi}, this effect is inverted for 
$m_{\phi}/m \approx 15$. As one raises the chemical potential this behavior is intensified,
as shown in Fig. \ref{3dgraf.ps}.

However, it is important to stress that, from a quantitative point of view, these results
depend strongly on the renormalization scale $\Lambda$ adopted. Although when analyzing a 
general theory it is not possible to discuss quantitatively the result, in the description of a specific
system the renormalization scale is determined through physical constraints, such as 
positivity of the energy density and the domain of energies being investigated.
In the case of QCD \cite{Fraga:2001id}, for example, it is not reasonable 
for quark matter to have a higher pressure than a hadron gas for low chemical potentials
(i.e., equivalent to densities of the order of the nuclear saturation density, $n_0\approx 0.16~\textrm{fm}^{-3}$),
since this would imply the existence of stable deconfined matter in a density regime
where only hadrons and nuclei are observed. Equivalently, one expects the quark pressure
to be higher than the hadronic one for densities $n\gg n_0$. In this way, one fixes a physically 
reasonable range for the renormalization scale in QCD. 
This kind of procedure can also be implemented once the phenomenology described 
by an effective theory containing a Yukawa sector is known, and the parameters are 
given by experimental observations. Therefore, once the physical picture is fixed, the predictions
above can be precised quantitatively.

\section{RG running effects}

The thermodynamic potential $\Omega_{Y}$ depends on the renormalization scale $\Lambda$
not only explicitly but also implicitly, through the RG running of the coupling and the masses, 
with $\Lambda$ corresponding to typical momenta involved in 
scattering processes in the medium \cite{footnote3}.
In the context of effective field theories, it is plausible 
to use the results above, with fixed $\alpha_Y$ and masses, and $\Lambda$ determined by 
phenomenological constraints. Nevertheless, for several applications, RG flow may be 
considered a relevant feature or even be intrinsically present, as in the case of sectors of 
the Standard Model. Then, one has to solve the corresponding RG equations and include 
their effects in the evaluation of the equation of state. In fact, these effects bring 
major consequences in cold and dense QCD \cite{Fraga:2001id,Fraga:2004gz}. Since 
we consider the general case of massive fermions and massive bosons, there will be 
an intricate competition between the different effects.
\begin{figure}[htb]
\vspace{0.5cm}
\begin{minipage}[t]{75mm}
\includegraphics[width=7.5cm]{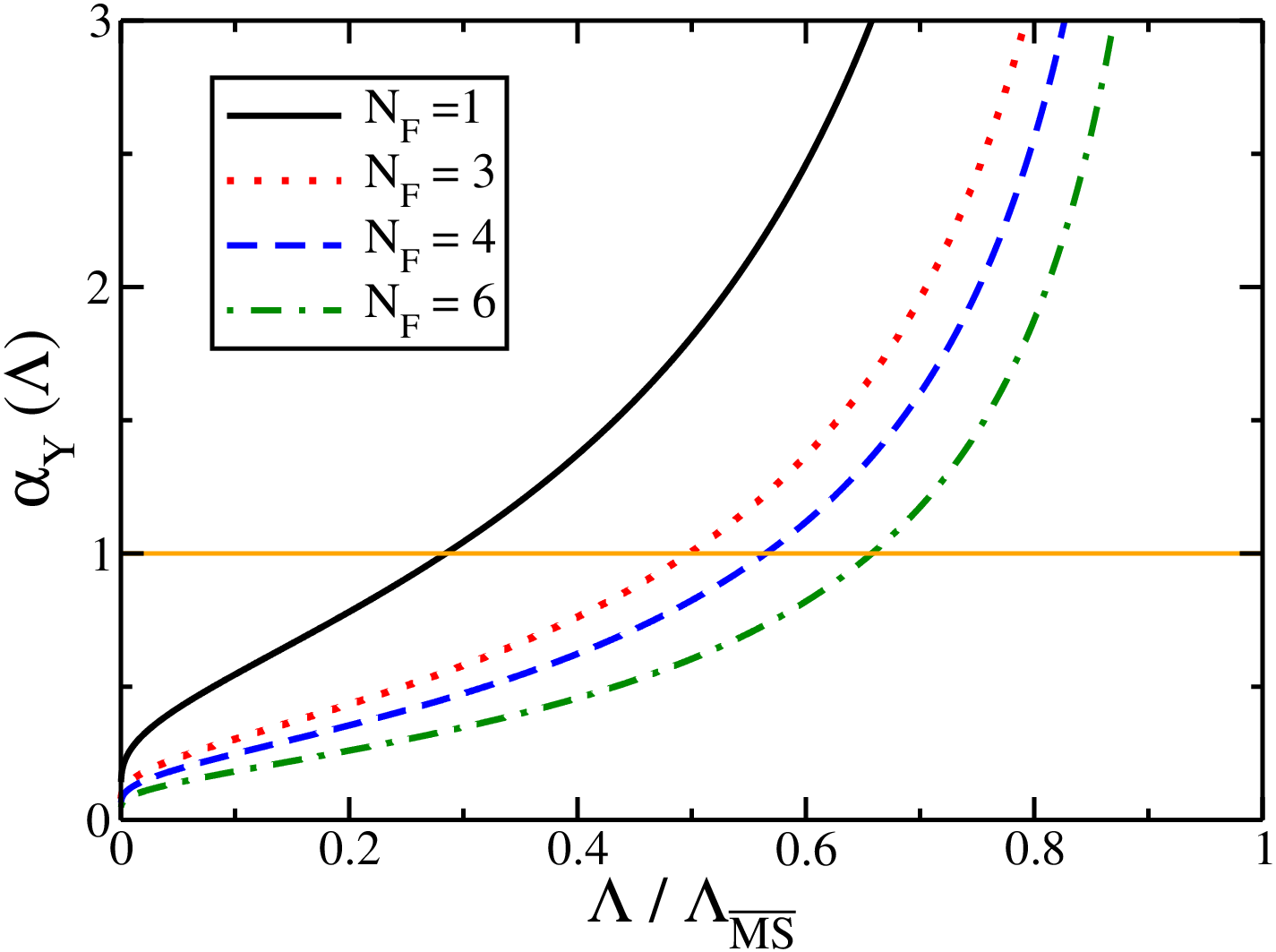}
\caption{Yukawa coupling RG flow normalized by the reference scale $\LMS$ for 
different numbers of fermion flavors.}
\label{runalphaY}
\end{minipage}
\hspace{2cm}
\begin{minipage}[t]{75mm}
%
\includegraphics[width=7.7cm]{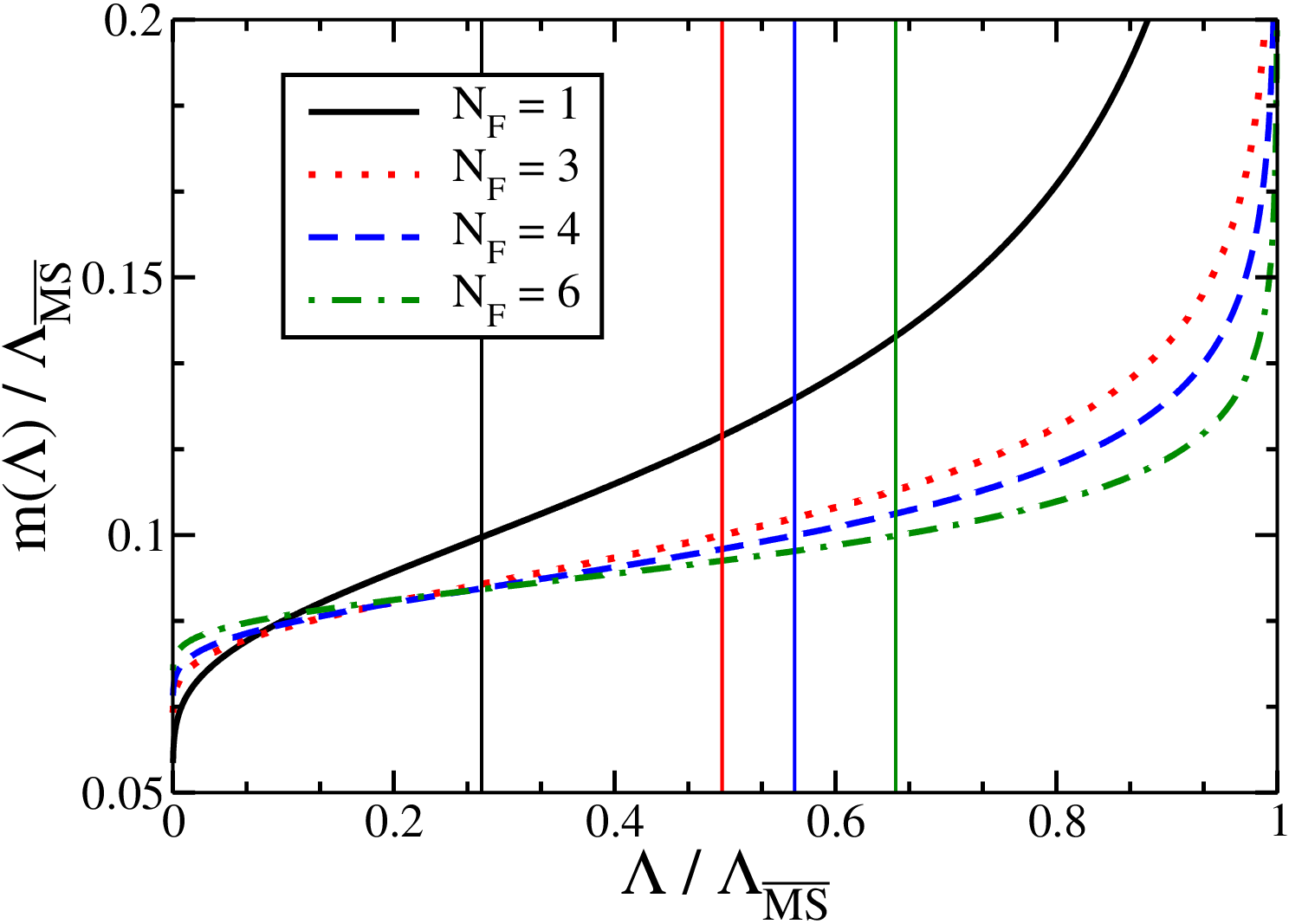}
\caption{Fermion mass RG flow normalized by the reference scale $\LMS$ for different numbers 
of fermion flavors. We chose $m_{0}=0.1\LMS$, and vertical lines represent the scale at which 
$\alpha_{Y}=1$ for each value of $N_{F}$, increasing from left to right.}
\label{runmass}
\end{minipage}
\end{figure}

In what follows we include the effects from running masses and coupling, and also 
investigate the role of the number of fermion flavors $N_F$. Although we study the behavior 
of RG running for different numbers of flavors, our results for the pressure are 
presented for $N_F = 4$. This is motivated by on-going studies on the lattice 
using Kogut-Susskind fermions \cite{taurines}.

The running coupling of the Yukawa theory up to $O(\alpha_Y)$ was computed long 
ago \cite{Coleman:1973sx} and reads
\be
\alpha_Y(\Lambda) &=& \frac{1}{\frac{3+2N_F}{2\pi}~
\log\left(\LMS/\Lambda \right)} 
\, . \label{Res-alphaY}
\ee
Fig. \ref{runalphaY} displays the scale dependence of the Yukawa coupling for 
different numbers of flavors. This  plot illustrates the crucial role of $N_F$ in the 
intensity of the interaction as a function of the energy scale and, therefore, in the 
delimitation of the domain of validity of perturbative calculations. Raising $N_F$ 
from $1$ to $6$, the energy scale at which the Yukawa coupling reaches one 
increases from $\sim 30\%$ to $\sim 70\%$ of the maximum scale $\LMS$, a very 
significant effect. This is also clear from Eq. (\ref{Res-alphaY}), where 
$\alpha_Y\sim 1/N_F$, which strongly suggests that a large $N_F$ approach could 
be appropriate for theories with Yukawa-type couplings as has been noticed 
long ago in a different context by Gross and Neveu \cite{Gross:1974jv}, who also 
identified the need to go beyond the large $N_F$ expansion in order to obtain 
sensible results (see also \cite{GN}).

The fermion mass runs according to \cite{ijmpe}
\be
m(\Lambda) &=& \left[ \alpha_Y(\Lambda) \right]^{\frac{3}{2(2N_F+3)}}~m_0
\, , \label{Res-mLambda}
\ee
where $m_0$ is the value of the mass at the scale where $\alpha_Y=1$, i.e.,
\be
m_0=m(\Lambda_0) \quad ; \quad \alpha_Y(\Lambda_0)=1
\, .
\ee
The RG flow of the fermion mass is plotted in Fig. \ref{runmass}. Once again, 
the number of flavors plays an important quantitative role. More interesting, 
though, is the following feature: in the large $N_F$ limit fermion masses
tend to become scale invariant, i.e., for larger values of $N_{F}$ the running 
of $m$ becomes flatter. In fact, from Eq. (\ref{Res-mLambda}), the $N_F\to \infty$ 
behavior of $m(\Lambda)$ is $\sim [(1/N_F)\log(\Lambda_0/\Lambda)]^{1/N_F}m_0$, 
rendering RG corrections to $m$ negligible very fast as $N_F$ increases. 
\vspace{0.3cm}
\begin{figure}[htb]
\vspace{0.5cm}
\begin{minipage}[t]{75mm}
\includegraphics[width=7.7cm]{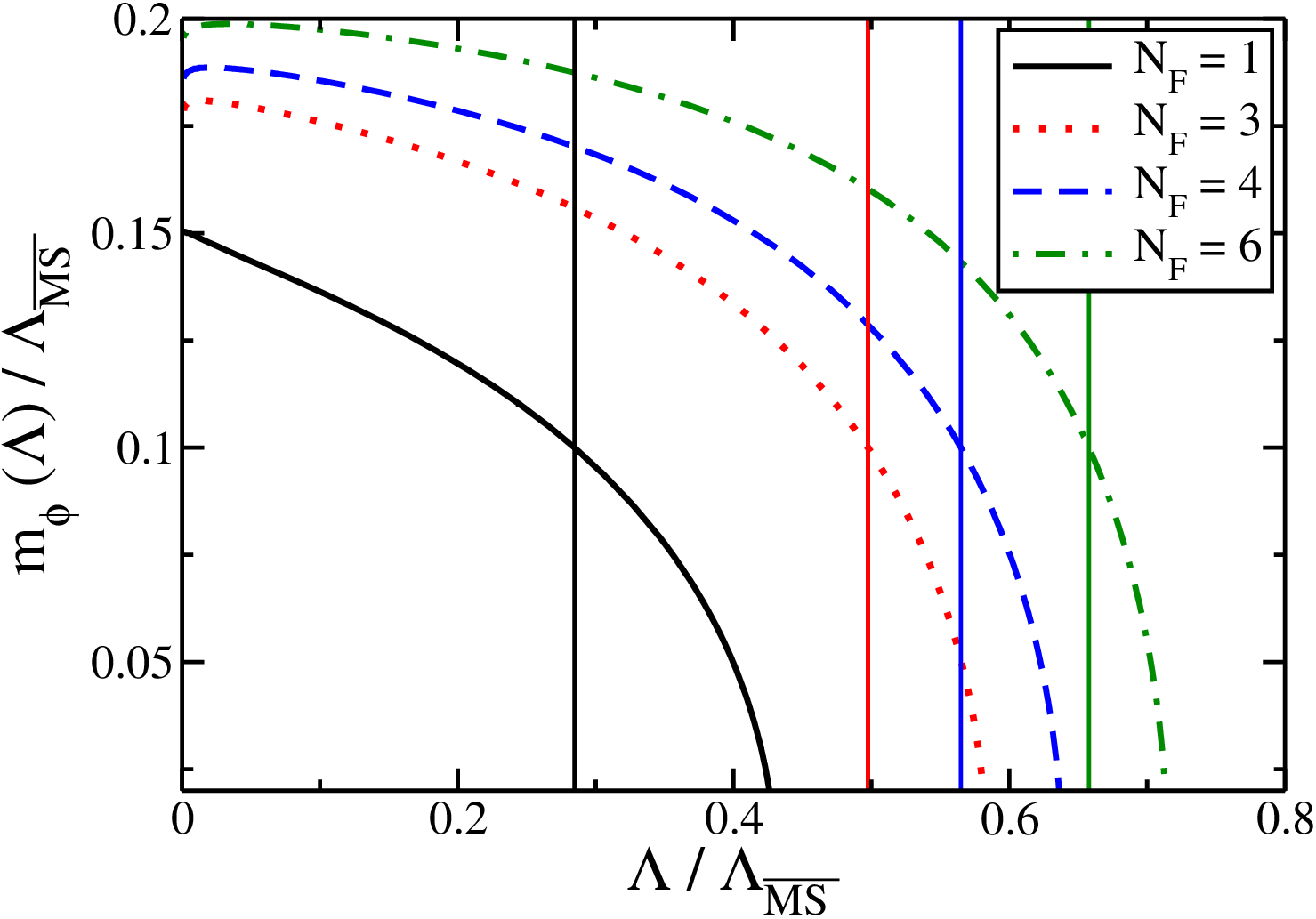}
\caption{Boson mass RG flow normalized by the reference scale $\LMS$ for different numbers 
of fermion flavors for $m_{0}=0.1\LMS$ and  $m_{\phi\, ,\, 0}=0.1\LMS$.}
\label{mphivsLambda-1}
\end{minipage}
\hspace{2cm}
\begin{minipage}[t]{75mm}
\includegraphics[width=7.7cm]{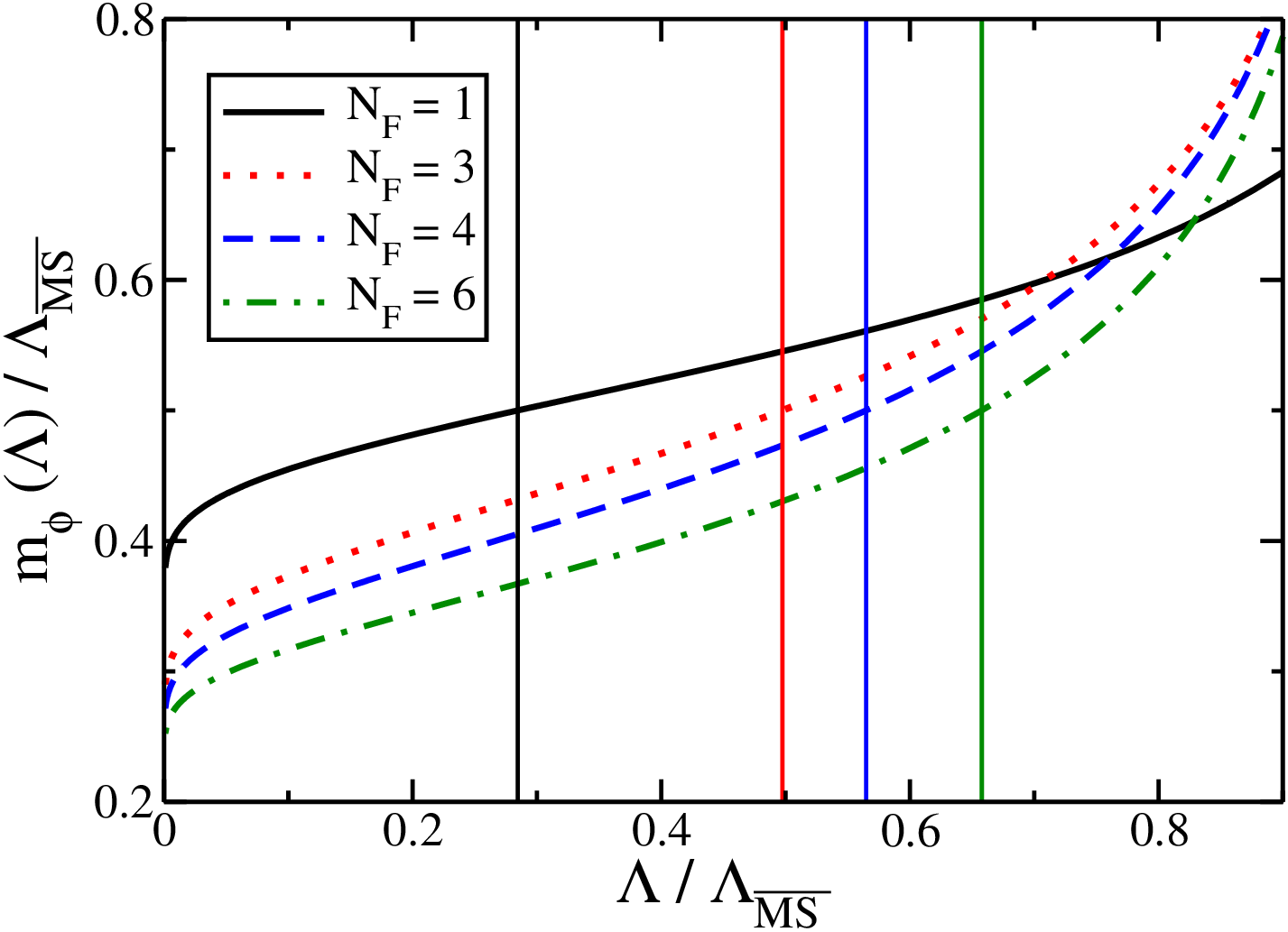}
\caption{Boson mass RG flow normalized by the reference scale $\LMS$ for different numbers 
of fermion flavors for $m_{0}=0.1\LMS$ and  $m_{\phi\, ,\, 0}=0.5\LMS$.}
\label{mphivsLambda-2}
\end{minipage}
\end{figure}

The scale dependence of the effective mass of the bosonic field is determined 
by the following flow equation
\be
\frac{\partial m_{\phi}^2}{\partial \log\left(\Lambda/\LMS\right)}
&=&\alpha_Y~\frac{N_F}{\pi}~[m_{\phi}^2-6m^2]
\label{EqRGmphi}
\, ,
\ee
with $\alpha_Y=\alpha_Y(\Lambda)$ and $m=m(\Lambda)$ given by (\ref{Res-alphaY}) and
(\ref{Res-mLambda}), respectively. The solution of this equation can be written in the 
following form:
\be
m_{\phi}^2(\Lambda)=m^2(\Lambda)\left\{ \frac{12N_F}{2N_F+3}+\mathcal{C}\left[ 
\log\left(\frac{\Lambda}{\LMS}\right) \right]^{-\,\frac{2N_F-3}{2N_F+3}} \right\}
\, ,\label{Res-mphi2Lambda}
\ee
where $\mathcal{C}$ is a constant fixed by the boundary condition 
$m_{\phi}(\Lambda_0)=m_{\phi , 0}$, and is given by
\be
\mathcal{C} &=&
\left[ 
\log\left(\frac{\Lambda_0}{\LMS}\right) \right]^{\frac{2N_F-3}{2N_F+3}}
~\left\{ \frac{m_{\phi\, , \, 0}^2}{m_0^2} - \frac{12N_F}{2N_F-3} \right\}
\, .
\ee
\begin{figure}[htb]
\vspace{-1.5cm}
\center
\includegraphics[width=12cm]{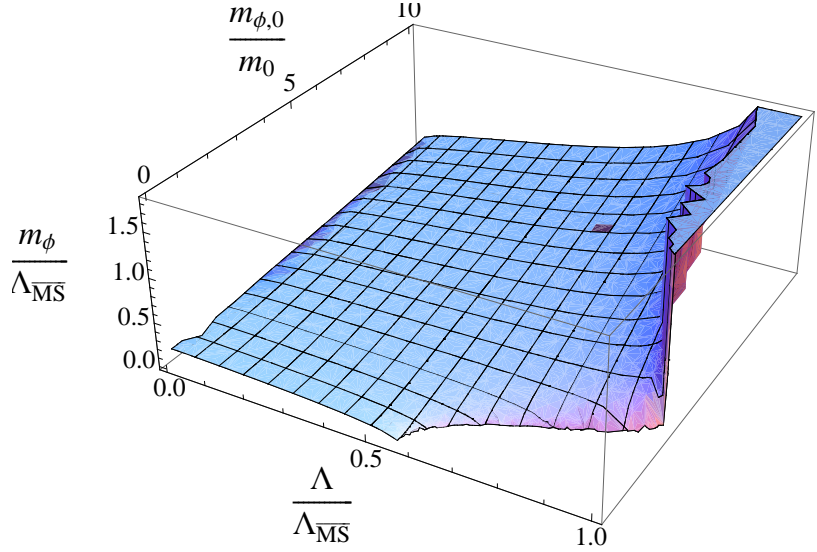}
\vspace{-7cm}
\caption{Running boson mass as a function of $\Lambda$ and the ratio 
of initial conditions $m_{\phi\, ,\, 0}/m_0$ in units of $\LMS$.}
\label{runmphi3D}
\end{figure}

It is clear from Eq. (\ref{EqRGmphi}) that the RG flow for $m_{\phi}$ exhibits two 
possible regimes, allowing the boson effective mass to increase or to decrease 
depending on the sign of the term inside the brackets. We illustrate these cases 
in Fig. 7 and Fig. 8 for different numbers of fermion flavors. We also show the 
interpolation between these two regimes in a three-dimensional plot, Fig. 9, as 
we vary the boundary conditions $m_{\phi\, , 0}/m_0$ for $N_F=4$. For large 
values of $N_{F}$, the boson mass is strongly modified (either increasing or 
decreasing smoothly, depending on the regime), and the RG flow does not seem 
to become negligible in either case.
\begin{figure}[htb]
\vspace{0.5cm}
\begin{minipage}[t]{75mm}
\includegraphics[width=7.7cm]{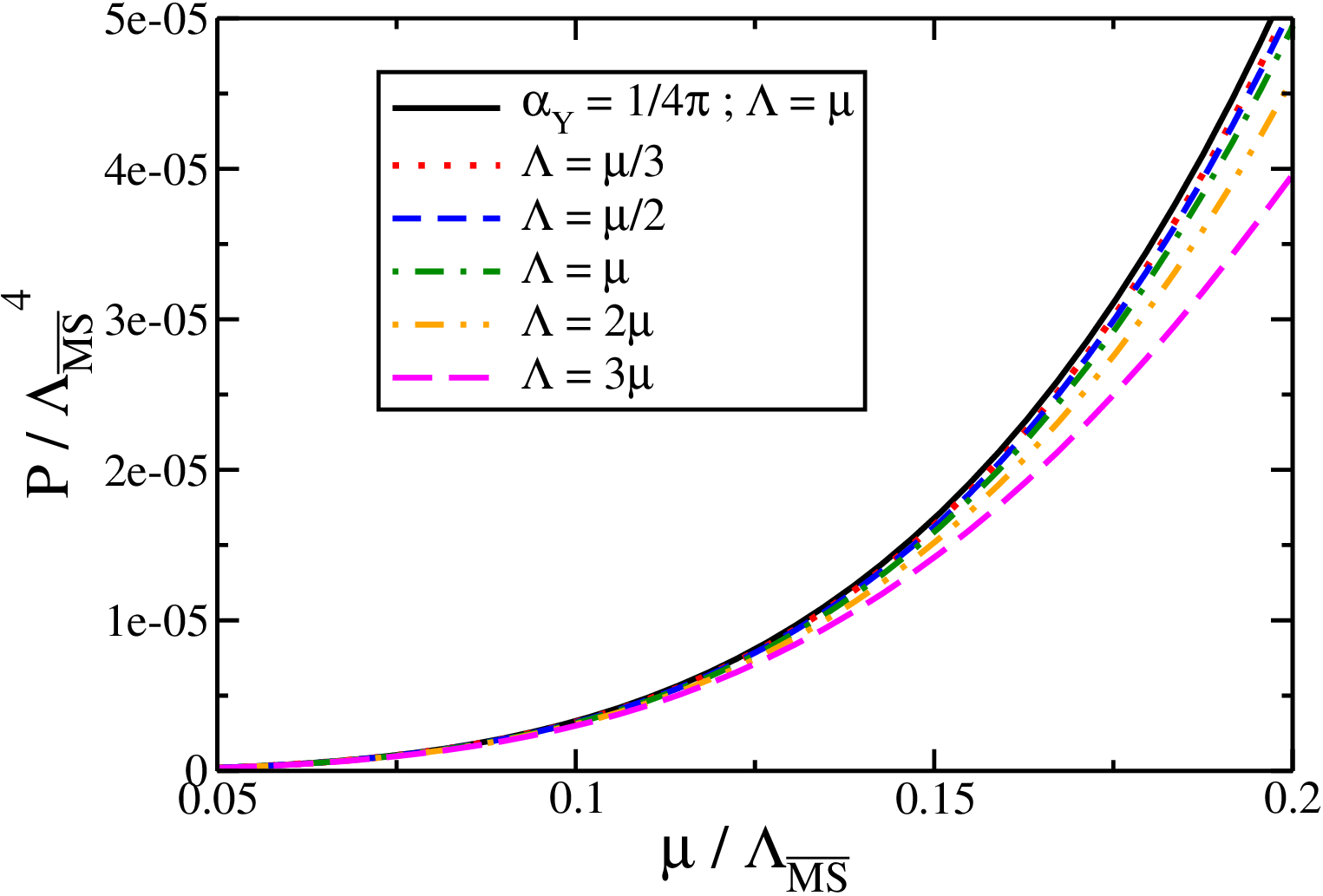}
\caption{Pressure versus chemical potential in units of the reference 
scale $\LMS$ for $N_F=4$ for different values of $\Lambda/\mu$ in the 
running of $\alpha_{Y}$. Here, $m=0$ and $m_{\phi}=0$.}
\label{RG-m0mphi0-Pvsmu}
\end{minipage}
\hspace{2cm}
\begin{minipage}[t]{75mm}
\includegraphics[width=7.7cm]{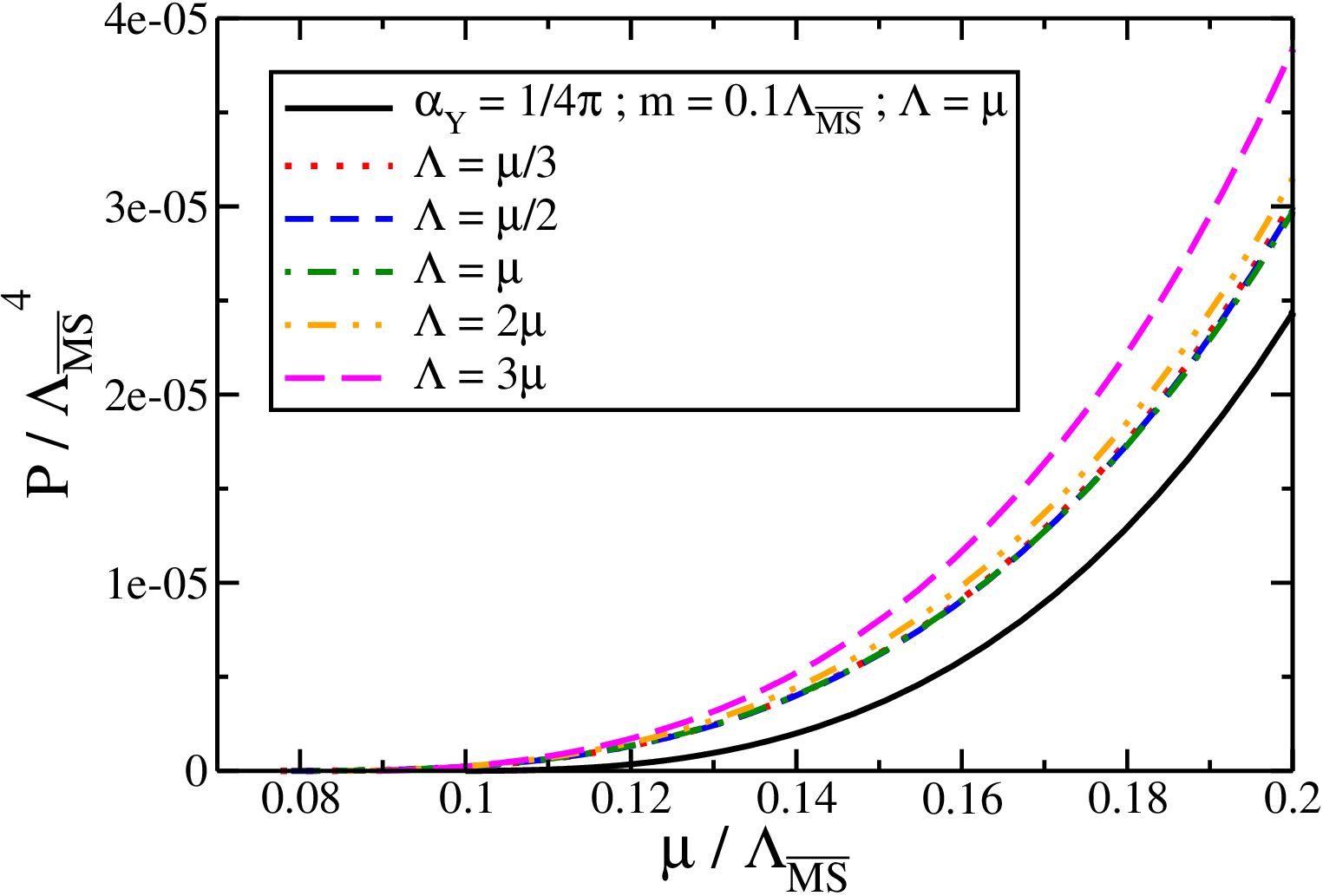}
\caption{Pressure versus chemical potential in units of the reference 
scale $\LMS$ for $N_F=4$ for different values of $\Lambda/\mu$ in the 
running of $\alpha_{Y}$ and $m$. Here, $m_{\phi}=0$ and $m_0=0.1 \LMS$.}
\label{RG-mphi0-Pvsmu}
\end{minipage}
\end{figure}

Since we consider a general Yukawa theory at finite density, the free parameters are arbitrary 
and allow for a wide spectrum of possibilities for the influence of the RG flow on the equation of 
state. So, although the RG flow is completely well defined by the results presented above, there 
still remains some freedom in the choice of the following scales: $\Lambda$, $\Lambda_0$, 
$m_0$ and $m_{\phi\, ,\, 0}$. The first usually represents the typical scattering energy scale, and 
its behavior should be motivated physically. In the case of cold and dense matter, the customary 
natural guess is simply $\Lambda\sim\mu$, the range of reasonable values for the proportionality 
constant being determined by physical consistency of the effective theory. The others must be 
fixed by measured observables. Therefore, all of them depend on the specificities of the system 
under consideration. In a given physical system, these parameters can be determined by 
phenomenological constraints, as discussed above, and no ambiguity is left. In the general 
theory we consider, which describes a whole class of possible physical systems, one 
can only exemplify the effects of the RG running by choosing a few illustrative cases.

RG flow effects on the pressure are displayed in Figs. 10 and 11 for certain choices 
of the free parameters, and for $N_{F}=4$. To simplify the analysis, we opted to keep 
the bosons massless. In Fig. 10, we show the pressure for different values of $\Lambda/\mu$ 
for massless fermions. It is clear that the pressure is lowered as one increases $\Lambda/\mu$.
In the case of massive fermions, displayed in Fig. 11, the effect of the RG flow is exactly the 
opposite. This apparent contradiction is not surprising. It is actually a natural consequence of 
the major influence of the fermion mass on the pressure.
The fact the the curve with fixed mass $m=0.1 \LMS$ is much lower than the rest is just an 
artifact of our choice of fixed mass being much bigger than the effective mass for these 
densities.

\section{Conclusions and outlook}

Yukawa theory at vanishing temperature provides at least one of the ingredients for 
an effective description of the thermodynamics of a variety of cold and dense 
fermionic systems, from condensed matter to particle physics. In this paper we 
have calculated the thermodynamic potential for a general Yukawa theory 
up to two loops within the $\overline{\textrm{MS}}$ scheme. We have explicitly 
considered the effect of {\it arbitrary} masses for both fields, the spinorial and the scalar, 
which renders the evaluation of two-loop integrals highly non-trivial. Nevertheless, all 
integrals were computed analytically and expressed in terms of well-known 
special functions. These results were verified against numerical results in several 
different regions of the parameter space. The role of RG running of coupling and 
masses was also investigated, as well as its dependence on the number of fermion 
flavors.

Our results show that the pressure is clearly lowered by increasing the mass of the fermions, 
this effect being much more relevant than the ones brought about by variations of the coupling. 
The effect of the scalar mass is also significant, and can be viewed, in a quasi-particle picture, 
as providing important corrections to the fermion mass which, as seen, modifies appreciably 
the equation of state. The role of the RG flow, on the other hand, is more fuzzy in the case of 
an arbitrary Yukawa theory, and our findings for its influence on the pressure were
illustrated in a few representative cases. Once the physical system is specified, though, these 
ambiguities are completely resolved. 

The behavior of running effects with increasing $N_{F}$ is also quite interesting for two reasons. 
First, it is rather convenient for perturbative purposes since $\alpha_Y$ remains smaller than one 
even for higher values of the energy scale. Then, one could, for instance, compare lattice results 
to perturbative calculations in a wider range of energy scale. Second, because the flow of the 
fermion mass tends to flatten out, becoming much simpler. Thus, our equation of state in the case of  
$N_F = 4$ can be useful, for instance, in on-going studies on the lattice using Kogut-Susskind 
fermions \cite{taurines} to deal with effective theories at finite density in the hope of bringing 
some understanding to the study of the Sign Problem \cite{Hands:2007by}. 

To study the phase structure of systems such as the antiferromagnetic/superconductor in the 
Hubbard model \cite{hubbard}, condensates in the core of neutron stars \cite{stars}, and QCD, 
among many others, one has to generalize our description and include the presence of 
a non-zero condensate by computing the full effective potential. This calculation is under 
way for the linear sigma model, and will be presented elsewhere \cite{LSM}.

\section*{Acknowledgments} 
The authors thank D. Rischke,  J. Schaffner-Bielich, and
A. Taurines for fruitful discussions. 
This work was partially supported by CAPES, CNPq, FAPERJ and FUJB/UFRJ.

\appendix*

\section{In-medium exchange diagram with massive fields}

In this appendix we discuss some technical details concerning the explicit
calculation of the in-medium exchange contribution to the thermodynamic potential
with both fermionic and bosonic non-zero masses.

\subsection{Matsubara sums}

From Eq. (\ref{excRF}), solving the trace over Dirac indices and using the following
representation of the Kronecker delta 
 \cite{Kapusta:1979fh}:
\be
\beta \delta_{n_1\, ,\, n_2+l} &=&
\frac{\textrm{e}^{\beta\left[
k^0+p_2^0-\mu\right]}-\textrm{e}^{\beta\left[
p_1^0-\mu\right]}}{p_1^0-p_2^0-k^0}
\, ,
\label{RepDelta}
\ee
one obtains

\vspace{-0.3cm}

%
%
\begin{eqnarray}
\parbox{10mm}{
\begin{fmfgraph*}(35,35)\fmfkeep{exchange}
\fmfpen{0.8thick}
\fmfleft{i} \fmfright{o}
\fmf{fermion,left,tension=.08}{i,o,i}
\fmf{dashes}{i,o}
\fmfdot{i,o}
\end{fmfgraph*}}
\quad &=& 
\beta V~g^2
\int \frac{d^3{\bf p}_1 d^3{\bf p}_2 d^3{\bf k}}{(2\pi)^6}
~\delta^{(3)}({\bf k}-{\bf p}_1+{\bf p}_2)
~\mathcal{S}({\bf p}_1,{\bf p}_2,{\bf k})
\, , \label{excRF3}
\end{eqnarray}
%
%
%

\noindent where
\be
\mathcal{S}({\bf p}_1,{\bf p}_2,{\bf k})&=&
T\sum_l
~T\sum_{n_ 1}~T\sum_{n_2}~\frac{s_0(p_1^0,p_2^0,k^0)}{(P_1^2-m^2)(m_{\phi}^2-K^2)(P_2^2-m^2)}
\, ,
\\ \nonumber \\
s_0(p_1^0,p_2^0,k^0)&=&
\frac{4 \,\left( P_1 . P_2 + m^2\right)}{p_1^0-p_2^0-k^0}
\left\{ \frac{1}{n_f(p_2^0-\mu)n_b(k^0)}-\frac{1}{n_b(k^0)}+\frac{1}{n_f(p_2^0-\mu)}
-\frac{1}{n_f(p_1^0-\mu)} \right\}
\, . \label{s0}
\ee

Resorting to the standard results
\be
T\sum_{l}\frac{g(k^0)}{m_{\phi}^2-K^2} &=&
\frac{1}{2\omega}\left.\left\{  g(k^0)~n_b(k^0) 
\right\}\right|_{k^0=-\omega}^{k^0=\omega}
\, ,
\\
T\sum_{n}\frac{g(p^0)}{P^2-m^2} &=& 
\frac{1}{2\Ep}\left.\left\{ g(p^0)~n_f(p^0-\mu) 
\right\}\right|_{p^0=-\Ep}^{p^0=\Ep}
\, ,
\ee
for the bosonic and fermionic Matsubara sums, respectively, 
with $g(q)$ an analytic function, the triple sum $\mathcal{S}$
reads:
\be
\mathcal{S}({\bf p}_1,{\bf p}_2,{\bf k})
&=&
\frac{1}{8\omega\Epu\Epd}\left.\Bigg\{
\frac{4 ~\left( P_1 . P_2 + m^2\right)}{p_1^0-p_2^0-k^0}
~\Big[ n_f(p_1^0-\mu)
-n_f(p_1^0-\mu)~n_f(p_2^0-\mu)+\right.
\nonumber \\ && \left.\left.\left.\quad
+n_f(p_1^0-\mu)~n_b(k^0)
-n_f(p_2^0-\mu)~n_b(k^0)
\Big]
\Bigg\}\right|_{p_1^0=-\Epu}^{p_1^0=\Epu}
\right|_{p_2^0=-\Epd}^{p_2^0=\Epd}
\right|_{k^0=-\omega}^{k^0=\omega}
\, .
\ee

After a straightforward algebraic manipulation using the properties of the Fermi-Dirac and 
Bose-Einstein distributions, the above equation can be written as
\be
\mathcal{S}({\bf p}_1,{\bf p}_2,{\bf k})
&=&
4~\frac{1}{8\omega\Epu\Epd}\Bigg\{
(m^2-{\bf p}_1\cdot {\bf p}_2+\Epu\Epd)~\left[ \frac{-2\omega}{E_-^2-\omega^2} \right]~
\Sigma_1
+\nonumber \\
&&
+(m^2-{\bf p}_1\cdot {\bf p}_2-\Epu\Epd)~\left[ \frac{2\omega}{-E_+^2+\omega^2} \right]~
\Sigma_2
+\nonumber \\
&&
+(m^2-{\bf p}_1\cdot {\bf p}_2+\Epu\Epd)~\left[ \frac{2E_-}{E_-^2-\omega^2} \right]~
n_b(\omega)~\left[ N_f(1)-N_f(2) \right]
+\nonumber \\
&&
+(m^2-{\bf p}_1\cdot {\bf p}_2-\Epu\Epd)~\left[ \frac{-2E_+}{E_+^2-\omega^2} \right]~
n_b(\omega)~\left[ N_f(1)+N_f(2) \right]
+\nonumber \\
&&
+\left[ \frac{m^2-{\bf p}_1\cdot {\bf p}_2+\Epu\Epd}{E_--\omega}-
\frac{m^2-{\bf p}_1\cdot {\bf p}_2-\Epu\Epd}{E_++\omega} \right]~N_f(1)
-\nonumber \\
&&
-\left[ \frac{m^2-{\bf p}_1\cdot {\bf p}_2+\Epu\Epd}{E_-+\omega}+
\frac{m^2-{\bf p}_1\cdot {\bf p}_2-\Epu\Epd}{E_++\omega} \right]~N_f(2)
+\nonumber \\
&&
+(m^2-{\bf p}_1\cdot {\bf p}_2-\Epu\Epd)~\left[ \frac{2E_+}{E_+^2-\omega^2} \right]~2n_b(\omega)
+\nonumber \\
&&
+2~\frac{m^2-{\bf p}_1\cdot {\bf p}_2-\Epu\Epd}{E_++\omega}
\Bigg\}
\, ,
\label{Sres}
\ee
where we used the definitions given in Eqs. (\ref{barJpm})--(\ref{Sig2}).

Finally, the expression in Eq. (\ref{excRes}) is obtained from the result (\ref{Sres})
by taking advantage of the symmetry of the integral in Eq. (\ref{excRF3}) to conveniently
exchange ${\bf p}_1 \leftrightarrow {\bf p}_2$ in the terms $\sim N_f(2)$.

\subsection{Renormalization}

Let us now sketch the proof that the UV divergences present
in the terms in Eq. (\ref{excRes}) that are linear in the statistical distributions,
\be
{\bf L}_f &\equiv&-\beta V~g^2
\int \frac{d^3{\bf p}_1 d^3{\bf p}_2}{(2\pi)^6}
~4~\frac{1}{8\omega_{12}\Epu\Epd}
\Big[ \overline{\mathcal{J}}_+ (E_-+\omega_{12})-\overline{\mathcal{J}}_- (E_+-\omega_{12}) \Big]
~N_f(1)
\, ,
\label{Lf}
\nonumber \\
{\bf L}_b &\equiv&\beta V~g^2
\int \frac{d^3{\bf p}_1 d^3{\bf p}_2}{(2\pi)^6}
~4~\frac{1}{8\omega_{12}\Epu\Epd}~
(-2)~\overline{\mathcal{J}}_-~E_+~n_b(\omega_{12})
\, , \label{Lb}
\ee
belong in fact to vacuum self-energy subdiagrams.

Using the auxiliary functions
\cite{footnote4}
%
\be
\mathcal{M}_f(p_1^4) &\equiv& \int_{-\infty}^{\infty}\frac{dp_2^4 dk^4}{(2\pi)^2}
~\frac{m^2+P_1\cdot P_2}{((p_2^4)^2+\Epd^2)((k^4)^2+\omega_{12}^2)}~2\pi 
\delta(k^4-p_1^4+p_2^4) \, ,
\label{Mf}
\\
\mathcal{M}_b (k^4)&\equiv&\int_{-\infty}^{\infty}\frac{dp_1^4 dp_2^4}{(2\pi)^2}
~\frac{m^2+P_1\cdot P_2}{((p_1^4)^2+\Epu^2)((p_2^4)^2+\Epd^2)}~2\pi 
\delta(k^4-p_1^4+p_2^4)
\, ,
\label{Mb}
\ee
where $P_i=(p_i^0,{\bf p_i})=(i p_i^4, {\bf p_i})$, which satisfy the 
identities:
\be
\mathcal{M}_f(p_1^4) &=& \frac{1}{4\Epd\omega_{12}} \left\{ 
\frac{m^2-{\bf p}_1\cdot{\bf p}_2-ip_1^4\Epd}{ip_1^4+\Epd+\omega_{12}}
-
\frac{m^2-{\bf p}_1\cdot{\bf p}_2+ip_1^4\Epd}{ip_1^4-\Epd-\omega_{12}}
\right\} \, ,
\label{Mf2}
\\
\mathcal{M}_b (k^4)&=& \frac{1}{4\Epu\Epd} 
(m^2-{\bf p}_1\cdot{\bf p}_2-\Epu\Epd) \left[ \frac{-2E_+}{E_+^2-(ik^4)^2} \right]
\, ,
\label{Mb2}
\ee
one can write:
\be
{\bf L}_f 
&=&-\beta V~g^2
\int \frac{d^3{\bf p}_1 d^3{\bf p}_2 d^3{\bf k}}{(2\pi)^6}~
\delta^{(3)}({\bf k}-{\bf p}_1+{\bf p}_2)
~4~\frac{2N_f(1)}{2\omega}~\mathcal{M}_f(-i\Epu) 
\, ,
\nonumber \\
{\bf L}_b 
&=&-\beta V~g^2
\int \frac{d^3{\bf p}_1 d^3{\bf p}_2 d^3{\bf k}}{(2\pi)^6}~
\delta^{(3)}({\bf k}-{\bf p}_1+{\bf p}_2)
~4~\frac{2n_b(\omega)}{2\omega}~
\mathcal{M}_b (-i\omega_{12})
\, ,
\ee
or, substituting the original expressions (\ref{Mf}) and (\ref{Mb}),
\be
{\bf L}_f 
&=&-\beta V~g^2~
\int \frac{d^3{\bf p}_1}{(2\pi)^3}~\frac{2N_f(1)}{2\omega}
\Bigg\{
\int \frac{d^3{\bf p}_2 d^3{\bf k}}{(2\pi)^6}~
\int_{-\infty}^{\infty}\frac{dp_2^4 dk^4}{(2\pi)^2}
~\times
\nonumber \\
&& \times
(2\pi)^3
\delta^{(3)}({\bf k}-{\bf p}_1+{\bf p}_2)
~2\pi 
\delta(k^4-p_1^4+p_2^4)
~\left.\frac{4(m^2+P_1\cdot P_2)}{((p_2^4)^2+\Epd^2)((k^4)^2+\omega_{12}^2)}
\Bigg\}\right|_{p_1^4=-i\Epu}
\, ,
\\
{\bf L}_b 
&=&-\beta V~g^2
\int \frac{d^3{\bf k}}{(2\pi)^3}
~\frac{2n_b(\omega)}{2\omega}~
\Bigg\{
\int \frac{d^3{\bf p}_1 d^3{\bf p}_2 }{(2\pi)^6}~
\int_{-\infty}^{\infty}\frac{dp_1^4 dp_2^4}{(2\pi)^2}
\times
\nonumber \\
&&\times~
(2\pi)^3
\delta^{(3)}({\bf k}-{\bf p}_1+{\bf p}_2)
~2\pi 
\delta(k^4-p_1^4+p_2^4)
\left.\frac{4(m^2+P_1\cdot P_2)}{((p_1^4)^2+\Epu^2)((p_2^4)^2+\Epd^2)}
\Bigg\}\right|_{k^4=-i\omega_{12}}
\, .
\ee

In Minkowski space, 
$P_i=(p_i^0=ip_i^4,{\bf p}_i)$, $K=(k^0=ik^4,{\bf k})$ and 
\be
{\bf L}_f 
&=&-\beta V~g^2~
\int \frac{d^3{\bf p}_1}{(2\pi)^3}~\frac{2N_f(1)}{2\omega}
\Bigg\{
\int \frac{d^4 P_2 d^4 K}{(2\pi)^4}~
(-i)
\delta^{(4)}(K-P_1+P_2)
~\left.\frac{4(m^2+P_1\cdot P_2)}{(-P_2^2+m^2)(-K^2+m_{\phi}^2)}
\Bigg\}\right|_{p_1^0=\Epu}
\, ,
\label{Lfsimp}
\ee
\be
{\bf L}_b 
&=&-\beta V~g^2
\int \frac{d^3{\bf k}}{(2\pi)^3}
~\frac{2n_b(\omega)}{2\omega}~
\Bigg\{
\int \frac{d^4 P_1 d^4 P_2 }{(2\pi)^4}~
(-i)
\delta^{(4)}(K-P_1+P_2)
\left.\frac{4(m^2+P_1\cdot P_2)}{(-P_1^2+m^2)(-P_2^2+m^2)}
\Bigg\}\right|_{k^0=\omega_{12}}
\, . \label{Lbsimp}
\ee

On the other hand, consider the one-loop diagrams contributing
to the amputated bosonic and fermionic vacuum self-energies.
When evaluated on the mass shell they satisfy:

\vspace{-0.3cm}

%
%
\begin{eqnarray}
\left(
\parbox{23mm}{
\fmfreuse{autoenergiaF}}
\right)_{\stackrel{amp}{m.s.}}^{vac}
&=&
\Bigg\{
g^2 \int \frac{d^4P_2d^4K}{(2\pi)^8}~(2\pi)^4
~\delta^{(4)}(K-P_1+P_2)~
\left.
\frac{2m^2+2P_1\cdot P_2}{
(P_2^2-m^2)(K^2-m_{\phi}^2)2m}
\Bigg\}\right|_{\slashed{P}_1=m}
\, , \label{autoenergiaF}
\end{eqnarray}
\begin{eqnarray}
\left(
\parbox{23mm}{
\fmfreuse{autoenergiaB}
}
\right)_{\stackrel{amp}{m.s.}}^{vac}
&=& 
\Bigg\{
-g^2 N_F\int \frac{d^4P_1d^4P_2}{(2\pi)^8}~(2\pi)^4
~\delta^{(4)}(K-P_1+P_2)~
\left.
\frac{4~(m^2+P_1\cdot P_2)}{(P_1^2-m^2)(P_2^2-m^2)}
\Bigg\}\right|_{K^2=m_{\phi}^2}
\, . \label{autoenergiaB}
\end{eqnarray}

Therefore, defining $P\equiv(i\omega_n^F+\mu,{\bf p}_1)$ and 
$Q\equiv(i\omega_l^B, {\bf k})$, we have:
\be
T\sum_{n}~\textrm{Tr}\left[\frac{1}{\slashed{P}-m}\right]
&=&\frac{N_F(1)-1}{2\Epu}~4m
\, ,
\\
T\sum_{l}~\frac{1}{m_{\phi}^2-Q^2} &=& \frac{2 n_b(\omega)+1}{2\omega}
\, ,
\ee
so that
\begin{eqnarray}
\sumint_P~(-1)\textrm{Tr}\left[\frac{1}{\slashed{P}-m}
\left(
\parbox{23mm}{
\fmfreuse{autoenergiaF}}
\right)_{\stackrel{amp}{m.s.}}^{vac}
\right]
&=&
-g^2
\int \frac{d^3{\bf p}_1}{(2\pi)^3}
\left[\frac{
N_f(1)-1}{2\Epu}\right]
\Bigg\{
\int \frac{d^4P_2d^4K}{(2\pi)^4}
~\delta^{(4)}(K-P_1+P_2)
\times
\nonumber \\
&&\quad\times
\left.\frac{4~(m^2+P_1\cdot P_2)}{
(P_2^2-m^2)(K^2-m_{\phi}^2)}
\Bigg\}\right|_{\slashed{P}_1=m}
\, , \label{"CTF2}
\end{eqnarray}
\begin{eqnarray}
\sumint_Q \frac{1}{m_{\phi}^2-Q^2}
\left(
\parbox{23mm}{
\fmfreuse{autoenergiaB}
}
\right)_{\stackrel{amp}{m.s.}}^{vac}
&=& -g^2 N_F \int \frac{d^3{\bf k}}{(2\pi)^3}\left[ \frac{
2n_b(\omega)+1
}{2\omega}
\right]
\Bigg\{
\int \frac{d^4P_1d^4P_2}{(2\pi)^4}
~\delta^{(4)}(K-P_1+P_2)\times
\nonumber \\
&&\left.\times~\frac{4~(m^2+P_1\cdot P_2)}{(P_1^2-m^2)(P_2^2-m^2)}
\Bigg\}\right|_{K^2=m_{\phi}^2}
\, . \label{"CTB2}
\end{eqnarray}

Comparing Eqs. (\ref{"CTF2}) and (\ref{"CTB2}) with Eqs.
(\ref{Lfsimp}) and (\ref{Lbsimp}), one concludes that ${\bf L}_f$ and ${\bf L}_b$ 
are written in terms
of the vacuum self-energies as stated in Eqs. (\ref{LfCT}) and (\ref{LbCT}).

Thus the UV renormalization of these terms results directly from adopting 
renormalized expressions for these self-energies. Using dimensional regularization 
in the $\overline{\textrm{MS}}$ scheme, the measure in momentum integrals 
is modified as
\be
\int\frac{d^4P}{(2\pi)^4} \mapsto \left(
\frac{\textrm{e}^{\gamma}\Lambda^2}{4\pi}
\right)^{\epsilon/2}\int\frac{d^d P}{(2\pi)^d}
\, ,
\ee
where $\epsilon=4-d$, with $d$ being the space-time dimension,
$\Lambda$ the renormalization scale and $\gamma$ the Euler constant.
A standard computation yields for the amputated vacuum one-loop self-energies
evaluated on the mass shell the following regularized expressions:

\begin{eqnarray}
\left(i~
\parbox{23mm}{
\fmfreuse{autoenergiaF}
}
\right)_{\stackrel{amp}{m.s.}}^{vac}
&=&
-\frac{g^2}{(4\pi)^2}
\int_0^1 dx~[m(1+x)]
~\left[ \frac{2}{\epsilon}+\log 
\left( \frac{\Lambda^2}{\Delta_f} \right) +O(\epsilon) \right]
\, , \label{autoenergiaFREGExp}
\end{eqnarray}
\begin{eqnarray}
\left(i~
\parbox{23mm}{
\fmfreuse{autoenergiaB}
}
\right)_{\stackrel{amp}{m.s.}}^{vac}
&=& 
\frac{g^2}{(4\pi)^2}N_F
\int_0^1 dx ~\Delta_b
~\left[ \frac{24}{\epsilon}+4+12\log 
\left( \frac{\Lambda^2}{\Delta_b} \right)
+O(\epsilon) \right]
\, , \label{autoenergiaBREGExp}
\end{eqnarray}
where
\be
\Delta_f &\equiv&  \left.\left\{ 
m^2(1-x)+m_{\phi}^2 x-x(1-x)P_1^2
\right\}\right|_{\slashed{P}_1=m} =m^2(1-x)^2+m_{\phi}^2 x
\, ,
\\
\Delta_b &\equiv&\left.\left\{m^2-x(1-x)K^2 \right\}\right|_{K^2=m_{\phi}^2}
=m^2-x(1-x)m_{\phi}^2
\, .
\ee
Defining the integrals over the Feynman parameter $x$ as
\be
\alpha_1&\equiv&\int_0^1 dx~(1+x)\log\left( \frac{\Lambda^2}{\Delta_f} \right)
\, ,
\\
\alpha_2&\equiv&\int_0^1 dx ~\Delta_b
\, ,
\\
\alpha_3&\equiv&\int_0^1 dx ~\Delta_b\log 
\left( \frac{\Lambda^2}{\Delta_b} \right)
\, ,
\ee
and substituting the renormalized expressions, obtained from Eqs. 
(\ref{autoenergiaFREGExp}) and (\ref{autoenergiaBREGExp})
by subtracting the poles in $\epsilon=0$,
in Eqs. (\ref{LfCT}) and (\ref{LbCT}), one arrives at the final results in Eqs. (\ref{LfCTRen}) 
and (\ref{LbCTRen}).
The solution of the integrals $\alpha_i$ is straightforward, yielding
Eqs. (\ref{alpha1res})--(\ref{alpha3res}).

\end{fmffile}
%
%
%

\subsection{The integral $J_1$}

Finally, in order to obtain the final analytical result in Eqs. (\ref{ResJ1final})--(\ref{ResJ1last}),
the integral in Eq. (\ref{defJ1}) must be calculated. Explicitly, we have:
\be
J_1 
&=&
\int \frac{d^3{\bf p}_1 d^3{\bf p}_2}{(2\pi)^6} \frac{\theta(\mu-E_1)\theta(\mu-E_2)}{2 E_1 E_2}
\left\{ \frac{4m^2-m_{\phi}^2}{(E_1-E_2)^2-| {\bf p}_1-{\bf p}_2 |^2-m_{\phi}^2}-1 \right\} \, ,
\\
&\equiv&
\frac{1}{(2\pi)^4}[j_I-2j_{II}]
\, ,\label{J1js}
\ee
where we used the expression for $\overline{\mathcal{J}}_{+}$ given in Eq. (\ref{barJpm})
and defined:
\be
j_I &\equiv& \int \frac{d^3{\bf p}_1 d^3{\bf p}_2}{(2\pi)^2} \frac{\theta(\mu-E_1)\theta(\mu-E_2)}{2 E_1 E_2}
 ~\frac{4m^2-m_{\phi}^2}{(E_1-E_2)^2-| {\bf p}_1-{\bf p}_2 |^2-m_{\phi}^2}
\, ,\\
j_{II} &\equiv&\frac{1}{2}\int \frac{d^3{\bf p}_1 d^3{\bf p}_2}{(2\pi)^4} \frac{\theta(\mu-E_1)\theta(\mu-E_2)}{2 E_1 E_2}
\, .
\ee
After evaluating the angular integration, one obtains:
\be
j_I&=& \frac{4m^2-m_{\phi}^2}{2}\int_m^{\mu}dE_1dE_2~\log \left[ 
\frac{2m^2-m_{\phi}^2-2E_1E_2+2p_1p_2}{2m^2-m_{\phi}^2-2E_1E_2-2p_1p_2} \right]
\equiv \frac{4m^2-m_{\phi}^2}{2}~\mathcal{I}_I
\nonumber\, ,\\
j_{II}&=& \int_m^{\mu}dE_1dE_2~p_1p_2 \equiv (\mathcal{I}_{II})^2
\, ,\label{jotinhas}
\ee
where $p_i\equiv \sqrt{E_i^2-m^2}$.

The integral $\mathcal{I}_{II}$ is simple, yielding:
\be
\mathcal{I}_{II} &=&\frac{1}{2}\left\{ \mu p_f-m^2\log\left[
\frac{\mu+p_f}{m} \right] \right\} \equiv \frac{u}{2}
\,.\label{I_IIRES}
\ee

On the other hand, the integral $\mathcal{I}_I$ is extremely more involved.
One of the dificulties is the fact that it corresponds to a double integral
of a $log$ whose argument contains roots which depend on the variable of integration.
It is possible, through a convenient transformation of variables, to obtain
an integrand without roots. A change of variables with this feature was presented
in Ref. \cite{Toimela:1984xy}:
\be
E_i &\mapsto& x_i = \frac{p_i}{E_i+m} \quad \Rightarrow\\
&& p_i = \frac{2mx_i}{1-x_i^2} \;~ ;~ \; E_i = m~\frac{1+x_i^2}{1-x_i^2} \; ~;~\; dE_i = \frac{4mx_i}{(1-x_i^2)^2}dx_i
\,.
\ee

Once this new set of variables is adopted, the procedure to solve the integral $\mathcal{I}_I$
includes a sequence of partial integrations and several algebraic manipulations until an expression
is reached such that it contains only analytically solvable integrals, at least for a 
software of algebraic computation.
Solving these integrals, one arrives at:
\be
\mathcal{I}_I &=& p_f^2 
-
2m^{2} ~\overline{\mathcal{K}}_{\log}\left( \frac{p_f}{\mu+m},\frac{m_{\phi}^2}{4m^2} \right)+
\mu^2~\frac{m_{\phi}^2}{2m^2} ~\log\left[ 
	\frac{m_{\phi}^2}{4p_f^2+m_{\phi}^2}
	\right]
	+\nonumber\\
	&&+
	\left(1-\frac{m_{\phi}^2}{2m^2}\right)\left\{ 2\mu p_{f}-m^2~
	\log\left[ \frac{\mu+p_f}{m} \right] \right\} ~
	\log\left[ \frac{m}{\mu+p_f} \right]
	+ \nonumber\\
&& 
+\frac{2m_{\phi}}{m} \sqrt{1-\frac{m_{\phi}^2}{4m^2}}
\left\{
\mu p_{f}-
\frac{m^{2}}{2}~
\log\left( \frac{\mu+p_f}{m} \right)
\right\}
\Bigg\{
\tan^{-1}\left(
\frac{-p_{f} m_{\phi}}{2m(\mu+m)\sqrt{1-\frac{m_{\phi}^2}{4m^2}}}
\right) + \nonumber\\
&&\quad+
\tan^{-1}\left(
\frac{p_{f}}{m_{\phi}\sqrt{1-\frac{m_{\phi}^{2}}{4m^{2}}}}
\left[2-\frac{m_{\phi}^2}{2m(\mu+m)} \right]
\right)
\Bigg\}
\,, \label{I7}
\ee
where $\overline{\mathcal{K}}_{\log}$ is the last integration still to be done:
\be
\overline{\mathcal{K}}_{\log}(\bar{x},z) &=& \sum_{\pm}~\sqrt{z(1-z)}~
\int_0^{\bar{x}}dx~\log\left( \frac{1-x}{1+x} \right) ~
\frac{d}{dx}\left\{ \tan^{-1}\left[ \frac{\frac{\bar{x}\pm x}{1-x^2}-\bar{x}~z}{\sqrt{z(1-z)}}
\right] \right\}
\,.
\ee

The solution of $\overline{\mathcal{K}}_{\log}$ can be written in terms of products of logarithms
whose arguments can assume negative values, resulting in a complex result in general.
These imaginary parts must cancel out in the end. To verify this cancellation and simplify the expression,
it is convenient to consider two complementary regimes: 
$z=m_{\phi}^2/4m^2~>~1$ and $z=m_{\phi}^2/4m^2~<~1$, and to use the fact that
$\bar{x}=p_f/(\mu+m)~<~1$. The final result, after a very long set of manipulations, 
is the one given in Eqs. (\ref{ResJ1final})--(\ref{ResJ1last}).



\begin{thebibliography}{9}
\bibitem{footnote1}
Up to two loop order, the diagrams with self-interaction vertices are either
one-particle reducible or independent of the fermionic chemical potential $\mu$.
\bibitem{footnote4}
The basic idea of this manipulation is to use the integral representation
of the Dirac delta function to factor the integrations and solve them independently,
as described in \cite{Romatschke-diplomathesis}.
\end{thebibliography}
\end{document}